\documentclass[aps,nofootinbib,twocolumn,superscriptaddress]{revtex4-1} 
\usepackage{graphicx}

\usepackage[english]{babel}
\usepackage{amssymb,amsfonts,psfrag,amsmath,bm,bbm,indentfirst,subfigure,bbold}
\usepackage{color}
\usepackage{ulem}
\usepackage{natbib}

\RequirePackage[
   hyperindex,colorlinks,bookmarksnumbered,
   plainpages=true,pdfstartview=FitH]{hyperref}
\hypersetup{linkcolor=blue,urlcolor=blue,citecolor=blue}
\usepackage{hyperref}
\usepackage{breakurl}

\definecolor{darkgreen}{rgb}{0,0.5,0}
\definecolor{purple}{rgb}{0.35,0,0.35}
\definecolor{orange}{rgb}{1,0.5,0}
\definecolor{darkred}{rgb}{.7,0,0}
\definecolor{darkblue}{rgb}{0.1,0.1,.6}
\definecolor{grey}{rgb}{.6,.6,.6}
\definecolor{dimgreen}{rgb}{0.2,0.6,0.1}
\definecolor{RoyalBlue}{cmyk}{0.94,0.539,0,0}
\definecolor{DGLorange}{cmyk}{.22,1,1,.2}
  
\newcommand{\Chi}{{\rm X}}
\newcommand{\bubble}{{W}}
\newcommand{\Eq}[1]{Eq.~(\ref{#1})}

\newcommand{\changed}[1]{{\color{black}{#1}}}

\newcommand{\pinchoff}{{\rm po}}   

\renewcommand{\emph}[1]{\textit{#1}}
\begin{document}
\allowdisplaybreaks

\title{Functional Renormalization Group Approach for Inhomogeneous
One-Dimensional Fermi Systems with Finite-Ranged Interactions}

\author{Lukas Weidinger}
\author{Florian Bauer}
\author{Jan von Delft}

\affiliation{Arnold Sommerfeld Center for Theoretical Physics and Center for
NanoScience,
Ludwig-Maximilians-Universit\"at
M\"unchen, Theresienstrasse 37, D-80333 M\"unchen, Germany}

\date{\today}

\begin{abstract}
We introduce an equilibrium formulation of the functional
renormalization group (fRG) for inhomogeneous systems capable of
dealing with spatially finite-ranged interactions.  In the general
third order truncated form of fRG, the dependence of the
two-particle vertex is described by $\mathcal{O}(N^4)$ independent
variables, where $N$ is the dimension of the single-particle
system. In a previous paper [Bauer~\textit{et al.}, Phys.~Rev.~B \textbf{89}, 045128
(2014)], the so-called coupled-ladder approximation (CLA) was
introduced and shown to admit a consistent treatment for models with
a purely onsite interaction, reducing the vertex to
$\mathcal{O}(N^2)$ independent variables.  In this work, we
introduce an extended version of this scheme, called the extended
coupled ladder approximation (eCLA), which includes a spatially
extended feedback between the individual channels, measured by a
feedback length $L$, using $\mathcal{O}(N^2 L^2)$ independent
variables for the vertex. \changed{We apply the eCLA in a static approximation and at zero temperature to three types of
one-dimensional model systems, focussing on obtaining the linear response conductance: First, we study a model of a quantum
point contact (QPC) with a parabolic barrier top and
onsite interactions. In our setup, where the characteristic length $l_x$ of the QPC ranges between approximately $4$-$10$ sites, eCLA achieves convergence
once $L$ becomes comparable to $l_x$.} It also turns out that the additional feedback stabilizes the
fRG-flow. This enables us, second, to study the geometric crossover
between a QPC and a quantum dot, again for a one-dimensional model
with \changed{onsite} interactions.  Third, the enlarged feedback also
enables the treatment of a finite-ranged interaction extending over
up to $L$ sites.  Using a simple estimate for the form of such
a finite-ranged interaction in a QPC with a parabolic barrier
top, we study its effects on the conductance and the
density.  We find that for low densities and sufficiently large
interaction ranges the conductance develops \changed{additional
 features}, and the corresponding density shows
some fluctuations that can be interpreted as Friedel oscillations
arising from a renormalized barrier shape with a rather flat top and
steep flanks.
\end{abstract}

\maketitle

\section{Introduction}
The functional renormalization group (fRG) is a well-established tool
for studying interacting many-body systems
\cite{Metzner2011,Andergassen2008,Karrasch2006a,Jakobs2010,Karrasch2006,Karrasch2010}.
This technique treats interactions using an RG-enhanced perturbation
theory and is known to provide an efficient way to treat
correlations. In particular, fRG can be used to treat spatially
inhomogeneous systems, represented by a discretized model with $N$
sites. For example, about $N\sim 10^2$ sites are required to represent
the electrostatic potential of a quasi-one-dimensional point contact
in a manner that is sufficiently smooth to avoid finite-sitze effects
\cite{Bauer2013}. The corresponding two-particle vertex has
$\mathcal{O}(N^4)\sim 10^8$ independent spatial components. To make
numerical computations feasible, simplifying approximations have to be
made to reduce the number of components used to describe the
vertex. Such a scheme, called the coupled-ladder approximation (CLA),
was proposed in Ref.~\onlinecite{Bauer2013} for the case of onsite
interactions.  Bauer, Heyder and von Delft (BHD) \cite{Bauer2014}
supplied a detailed description of the CLA which is in principle
applicable to systems of arbitrary dimensionality. The CLA is
implemented within the context of generic, third-order-truncated fRG,
meaning that all vertices with three and higher particle number are
set to zero throughout the whole flow.  In this paper we generalize
this scheme to be able to treat finite-ranged interactions. Since the
central aim of our scheme is to extend the spatial range over which
information is fed back into the RG flow, we call our scheme the
\textit{extended coupled-ladder approximation} (eCLA).

The basic idea of the CLA, and by extension the eCLA, lies in reducing
the number independent components of the vertex by decomposing it into
several interaction channels and then establishing a consistent
approximation by controlling the amount of feedback between the
individual channels. This strategy follows that used in
Refs.~\onlinecite{Karrasch2008,Jakobs2010} in the context of the
single-impurity Anderson model. For a model with short-ranged
interactions, this approach reduces the number of independent
quantities in the vertex to order $\sim \mathcal{O}(N^2)$.  From a
perturbative point of view, this treatment is exact in second order in
the interaction and amounts to summing up \changed{approximate contributions from a large class of
diagrams}, including mutual feedback between the different
interaction channels.  The eCLA generalizes the CLA by extending
spatial feedback between the channels. As a control parameter for this
extended feedback we introduce a feedback length $L$, where $L=0$
corresponds to the previous approximation scheme used by BHD, while
$L=N-1$ includes the full fRG flow in second order. $L$ thus serves as
a control parameter for the number of independent spatial components
of the vertex, which scales as $\sim \mathcal{O}(N^2L^2)$.  Moreover,
the longer-ranged feedback allows us also to treat interactions with
finite range up to $L_U$ sites (with $L_U\leq L$) in a manner that is
exact to second order in the interaction.

In this paper, we present a detailed account of the eCLA, and apply it to two one-dimensional (1D) fermionic systems, modeled to describe the lowest 1D subband of a quantum point contact (QPC) or a quantum dot (QD), respectively.
We develop the eCLA for systems described by a Hamiltonian of the form
\begin{align}
 \hat{H}=\sum_{ij, \sigma}  h_{ij}^{\sigma} d^\dagger_{i\sigma} d_{j\sigma}^{} +
 \frac{1}{2}\sum_{ij,\sigma\sigma'}U_{ij} \hat{n}_{i\sigma} \hat{n}_{j\sigma'} 
 (1-\delta_{ij}\delta_{\sigma \sigma'}) \; ,
 \label{eqModelGeneral}
\end{align}
where $h^\sigma$ and $U$ are real, symmetric matrices,
$d^\dagger_{j\sigma}$ creates an electron in single particle state $j$
with spin $\sigma$ ($=\uparrow,\downarrow$ or $+,-$, with
$\bar{\sigma}\!=\!-\sigma$), and
$n_{j\sigma}= d^\dagger_{j\sigma} d_{j\sigma}^{}$. In the context of the
applications presented here, we refer to the quantum number $j$ as the
``site index''.  Our eCLA scheme requires the interaction to have a
finite range $L_U\leq L$, such that
\begin{align}\label{eqCondU}
 U_{ij} = 0 \quad {\rm if} \; |i-j|>L_U.
\end{align}
Models of this form, but with onsite interactions
$(U_{ij} = U \delta_{ij}$), have been used to study both QPC and QD
systems \cite{Bauer2013}.  To describe a QPC, $h^\sigma_{ij}$ is taken
to represent a one-dimensional tight-binding chain, with a potential
barrier with parabolic top, whereas for a QD it is chosen to represent
a double-barrier potential. The non-interacting physics of both models
is well known, whereas the effect of interactions, especially for the
QPC, are still a topic of ongoing discussions
\cite{Meir2008,Sloggett2008,Aryanpour2009}. For the QPC the
conductance is quantized \cite{Wees1988,Wharam1988,Buttiker1990} in
units of the conductance quantum $G_Q={2e^2}/{h}$, but shows an
additional shoulder at approximately $0.7 G_Q$. This regime, in which
other observables show anomalous behavior too
\cite{Thomas1996,Appleyard2000,Cronenwett2002}, is commonly known as
the ``0.7-anomaly''.  The latter has been studied in \cite{Bauer2013}
using a model of the above form, with purely onsite
interactions. However, to examine the effect of gate-induced screening
in a QPC, one needs to consider finite-ranged interactions. This goal
serves as the main motivation for developing the eCLA put forth in
this paper. 

\changed{We remark that the QD and QPC models considered
  here provide a meaningful testing ground for the eCLA, since
  lowest-order perturbation theory would not yield an adequate
  treatment of the correlation effects expected to occur: the Kondo
  effect for QDs and the 0.7-anomaly for QPCs. Although some aspects
  of the latter can be understood in terms of a simple Hartree picture
  \cite{Bauer2013}, the interaction strength  needed to 
  yield phenomenological behavior typical of the 0.7-anomaly is
  sufficiently large that lowest-order perturbation theory is inadequate.}

The numerical results presented here were all obtained using the eCLA
in a static approximation, which neglects the frequency-dependence of
the two-particle vertex (after which the approach no longer is exact
to second order). Nevertheless, BHD have shown that for a QPC model
with onsite interactions, the CLA with a static approximation leads to
reasonable results for the conductance step shape, though it does
produce some artifacts regarding the pinch-off gate voltage when the
interaction strength is increased. We find the same to be true for the
static eCLA, with the artifacts becoming more pronounced with
increasing interaction range, but the step shape behaving in a
physically reasonable manner.

We use the eCLA for three studies of increasing complexity. (i) We
present static eCLA results for a QPC model with short-ranged
interaction and successively increase the feedback length $L$. This
systematically improves the treatment of RG-feedback between the
various fRG channels, and for sufficiently large $L$ converges to the
full solution of the generic, third-order-truncated static fRG. 
\changed{For the models we consider here, where the characteristic length $l_x$ of the parabolic QPC potential barrier
varies between approximately $4-10$ sites, we find that convergence in $L$ is achieved once $L$ becomes comparable to $l_x$.}
For such systems, the eCLA scheme thus speeds up the calculation relative to the full
generic, third-order-truncated static fRG by a factor of $10^3$,
without any loss of accuracy.  (ii) Furthermore, it turns out that the
eCLA's enhanced feedback leads to a more stable fRG flow than the
CLA scheme, since each interaction channel acts more strongly to
limit the tendencies other channels might have to diverge during the
fRG flow.  This enables us to study the geometric crossover between
a QPC and a QD where the barrier top stays close to the chemical
potential. This setup features a high local density of states (LDOS)
at the chemical potential, and as a result turns out to be intractable
when using the CLA without enhanced feedback \cite{Heyder2015}.  In
contrast, the eCLA is able to treat this challenging crossover very
nicely.  (iii) Finally, we illustrate the potential of the eCLA
to deal with finite-ranged interactions in a setting where the
physics of screening comes into play, namely a QPC model with an
interaction whose range extends over up to $N$ sites. The purpose of
this study is mainly methodological, i.e.\ we do not aim here to
achieve a fully realistic treatment of screening in a
QPC. Nevertheless, the results are interesting: for a sufficiently
long ranged interaction and sufficiently low density, there exists a
parameter regime where we find \changed{additional features in the conductance}
 and corresponding $2k_F$ density fluctuations \emph{within}
the QPC.

The paper has three main parts. The first part
(section~\ref{secfRG}) develops our improved eCLA feedback scheme.
The second part (section~\ref{secfRGresults}) studies its
consequences for QPC and QD models with onsite interaction, focusing
on the effects of increasing the feedback length $L$.  Finally, the
third part (section ~\ref{secFiniteInt}) is devoted to finite-ranged
interactions. We estimate the approximate form and strength of the
interaction to be used for a 1D depiction of a QPC and show some
preliminary results for the conductance and density profile of such
a system depending on the screening properties. A detailed study of
the physics of long-ranged interactions in QPCs is beyond the scope
of this work and left as a topic of future investigation.

\section{fRG flow equations}\label{secfRG}
Before we introduce our new eCLA scheme, we give a short overview over the general idea and the usual approximations made in fRG.
Since numerous detailed treatments of fRG are available, and since our work builds on that of BHD, the discussion below is very brief and structured similarly to that in Ref.~\onlinecite{Bauer2014}.
The basic idea of fRG is to introduce a flow parameter $\Lambda$ in the bare propagator of the theory in such a way that for $\Lambda=\Lambda_i=\infty$, the structure of the resulting vertex functions are very simple. With our choice for $\Lambda$ (described later) all but the two-particle vertex will vanish,
\begin{eqnarray}
\label{eq:initial-condition-vertex}
\gamma^{\Lambda_i}_2=v \quad \gamma^{\Lambda_i}_n=0 \quad (n \neq 2) ,
\end{eqnarray}
where $v$ is the bare vertex. For the final value of the flow parameter $\Lambda=\Lambda_f=0$, one recovers the full bare propagator and hence the full theory:
\begin{equation}
  \mathcal{G}_0^\Lambda \rightarrow \mathcal{G}_0 \; , 
  \quad \textrm{with} \quad
  \mathcal{G}_0^{\Lambda_i}=0, \quad \mathcal{G}_0^{\Lambda_f}=\mathcal{G}_0 \; .
\end{equation} 
The RG flow is described by a hierarchy of coupled differential equations for the one particle irreducible (1PI) $n$-particle vertex functions $\gamma_n$,
\begin{align} \frac{d}{d \Lambda} \gamma^\Lambda_n = \mathcal{F}
  \left( \Lambda,\mathcal{G}^{\Lambda}_0,\gamma^\Lambda_1, \dots
    ,\gamma^\Lambda_{n+1} \right) \;
  .  \label{eq:general-flow-equations}
\end{align}
Integrating this system from $\Lambda=\Lambda_i$ to $\Lambda=0$ yields
in principle a full description of all interaction vertices.  In
practice, one can of course not treat an infinite hierarchy of flow
equations and has to truncate it at some point. In our form of
third-order truncated fRG, we incorporate the one- and two-particle
vertex into the flow, but set all vertices with three or more
particles to zero
\begin{equation}
\label{eqfRGtrunc}
 \frac{d}{d \Lambda} \gamma_n = 0 \qquad (n \geq 3) \,.
\end{equation}
We thus retain only the flow of the self-energy, $\Sigma=-\gamma_1$, and the flow of the two-particle vertex $\gamma_2$. This differential equation can then be solved numerically, using a standard Runge-Kutta method.
As we will see shortly, the flow of the vertex consists of three different parquet-like channels which are coupled to the flow of the self-energy and also directly to each other. This simultaneous treatment moderates competing instabilities in an unbiased way.

In principle, the form of the fRG flow equations depends on the choice of the flow parameter, even if in most cases they take the form stated below. In our work, we choose the $\Lambda$-dependence of the bare propagator to take the form of an infrared cutoff
\begin{equation}
\label{eqfRGscheme}
\mathcal{G}_0^{\Lambda} (\omega_n)
 =\Theta_T (| \omega_n | - \Lambda ) \mathcal{G}_0 (\omega_n) \; ,
\; \Lambda_i = \infty, \; \Lambda_f = 0 \; .
\end{equation}
We use the Matsubara formalism with the frequencies $\omega_n$ defined to be purely imaginary,
\begin{equation*}
 \omega_n=iT\pi(2n+1) ,
\end{equation*}
and $\Theta_T$ is a step function broadened on the scale of temperature.

Using this cutoff, one can derive the fRG equations in the standard way, see e.g. Refs.~\onlinecite{Karrasch2006,Bauer2014a} or Ref.~\onlinecite{Jakobs2007} for a diagrammatic derivation. The resulting equation for the one-particle vertex is given by 

\begin{align}
\label{eqgamma1DGL}
\frac{\textrm{d}}{\textrm{d} \Lambda} \gamma_1^\Lambda
( \color{DGLorange} q_1' \color{black}, \color{RoyalBlue} q^{~}_1
 \color{black}) =
 T \sum_{q_2',q_2^{~}}
 \mathcal{S}_{q^{~}_2,q'_2}^\Lambda
 \gamma_{2}^\Lambda
(q_2' ,\color{DGLorange} q_1' \color{black}; q^{~}_2,
\color{RoyalBlue} q^{~}_1  \color{black}) \;  ,
\end{align}
where $q_i$ is a shorthand for all quantum numbers and the
fermionic Matsubara frequency associated with the legs of a vertex, and the full- and single-scale propagators are defined via
\begin{subequations}
\label{eqfRGgreen}
\begin{align}
\label{eqFullPropergator}
& \mathcal{G}^\Lambda =
\Big[ \left[ \mathcal{G}^{\Lambda}_0 \right]^{-1} - \Sigma^{\Lambda}
\Big]^{-1}
\; , \\
& \mathcal{S}^\Lambda = \mathcal{G}^\Lambda
\partial_\Lambda \left[ \mathcal{G}^{\Lambda}_0 \right]^{-1} \mathcal{G}^\Lambda,
\; 
\end{align}
\end{subequations}
respectively. The structure of the vertex consists naturally of three different parquet-like channels 
\begin{equation}
\label{eq:Ansatz-vertex-four-parts}
\gamma_2^\Lambda  = v + \gamma_p^\Lambda + \gamma_x^\Lambda + \gamma_d^\Lambda \, ,
\end{equation}
where $v$ is the bare vertex and we refer to  $\gamma_p^\Lambda$, $\gamma_x^\Lambda$, and
$\gamma_d^\Lambda$ as the particle-particle channel ($P$), and the exchange ($X$) and direct ($D$)
part of the particle-hole channel.
These quantities are defined via their flow equations 
\begin{align}
\phantom{.} \hspace{-0.7cm} \frac{\textrm{d}}{\textrm{d}
\Lambda}  \gamma_2^\Lambda
&  = \frac{\textrm{d}}{\textrm{d} \Lambda} ( \gamma_p^\Lambda
+
\gamma_x^\Lambda 
+
\gamma_d^\Lambda ) \; , 
\label{eq:diagrammatic-flow-equations}
\end{align}
and the initial conditions  $\gamma_p^{\Lambda_i}=\gamma_x^{\Lambda_i}=\gamma_d^{\Lambda_i}=0$.
The explicit form of the flow equations is
\begin{widetext}
\begin{subequations}
\label{eqgamma2DGL}
\begin{align}
\frac{\textrm{d}}{\textrm{d} \Lambda}  \gamma_p^\Lambda
(\color{DGLorange} q_1',q_2' \color{black} ;
\color{RoyalBlue} q^{~}_1,q^{~}_2 \color{black}) & =
\phantom{-}
T \sum_{q_3', q^{~}_3, q_4', q^{~}_4}
\hspace{-3mm}\gamma_2^\Lambda
(\color{DGLorange} q_1',q_2' \color{black} ;q^{~}_3,q^{~}_4)
\mathcal{S}^\Lambda_{q^{~}_3, q_3'}
\mathcal{G}^\Lambda_{q^{~}_4,q_4'}
\gamma_2^\Lambda (q_3', q_4';
\color{RoyalBlue} q^{~}_1,q^{~}_2 \color{black})  ,
\label{eqgamma2DGL_P}
\\
\frac{\textrm{d}}{\textrm{d} \Lambda}  \gamma_x^\Lambda
(\color{DGLorange} q_1',q_2' \color{black} ;
\color{RoyalBlue} q^{~}_1,q^{~}_2 \color{black}) & =
\phantom{-}
T \sum_{q_3', q^{~}_3, q_4', q^{~}_4}
\gamma_2^\Lambda
(\color{DGLorange} q_1' \color{black} ,q_4';
q^{~}_3, \color{RoyalBlue} q^{~}_2 \color{black})
\Bigl[\mathcal{S}^\Lambda_{q^{~}_3,q_3'}
\mathcal{G}^\Lambda_{q^{~}_4, q_4'}  
+\mathcal{G}^\Lambda_{q^{~}_3,q_3'}
\mathcal{S}^\Lambda_{q^{~}_4, q_4'}  \Bigr]
\gamma_2^\Lambda (q_3', \color{DGLorange} q_2' \color{black} ;
\color{RoyalBlue} q^{~}_1 \color{black} , q^{~}_4)\; ,
\label{eqgamma2DGL_X}
\\
\frac{\textrm{d}}{\textrm{d} \Lambda}  \gamma_d^\Lambda
(\color{DGLorange} q_1',q_2' \color{black} ;
\color{RoyalBlue} q^{~}_1,q^{~}_2 \color{black}) & =
-
T \sum_{q_3', q^{~}_3, q_4', q^{~}_4}
\gamma_2^\Lambda (\color{DGLorange} q_1' \color{black} ,q_3';
\color{RoyalBlue} q^{~}_1 \color{black} , q^{~}_4)
\Bigl[\mathcal{S}^\Lambda_{q^{~}_4,q_4'}
\mathcal{G}^\Lambda_{q^{~}_3, q_3'}  
+\mathcal{G}^\Lambda_{q^{~}_4,q_4'}
\mathcal{S}^\Lambda_{q^{~}_3, q_3'}  \Bigr]
\gamma_2^\Lambda (q_4', \color{DGLorange} q_2' \color{black} ;
q^{~}_3, \color{RoyalBlue} q^{~}_2 \color{black}) \; .
 \label{eqgamma2DGL_D}
 \end{align}
\end{subequations}
\end{widetext}
At this point, the channels have a full feedback between them. Later on, however, we will control the amount of feedback between channels by the feedback length $L$.

\subsection{Frequency Parametrisation}
Since we have energy conservation at each vertex,
\begin{equation}
\begin{array}{l}
 \vspace{2mm}
 \gamma_1(q_1',q_1^{~}) \propto \delta_{n_1'n_1^{~}},\\ 
 \gamma_2 (q_1',q_2';q_1^{~},q_2^{~})  \propto \delta_{n_1'\!+\!n_2'n_1^{~}\!+n_2^{~}},
\end{array}
\end{equation}
we can parametrize the frequency dependence of the self-energy with one frequency, and of the vertex with three frequencies. A detailed discussion of the frequency structure is given in Refs.~\onlinecite{Karrasch2008,Jakobs2010,Bauer2014}, and since we proceed analogously, we will be very brief here. A convenient choice for the parametrization of the vertex frequency structure is given in terms of the three bosonic frequencies \cite{Bauer2013}
\begin{subequations}
\begin{align}
 \Pi =&\, \omega_{n_1'} + \omega_{n_2'} = \omega_{n_1^{~}} + \omega_{n_2^{~}} \, ,\\
 \Chi =&\,\omega_{n_1'}-\omega_{n_2^{~}} = \omega_{n_1^{~}} - \omega_{n_2'}  \, , \\
 \Delta =&\, \omega_{n_1'} - \omega_{n_1^{~}} = \omega_{n_2^{~}} - \omega_{n_2'} \, .
\end{align}
\end{subequations}
In order to keep notation short, the frequency information is separated from the site and
spin quantum numbers:
\begin{flalign}
& \gamma_2(j_1' \sigma_1' \omega_{n_1'}, j_2' \sigma_2' \omega_{n_2'};
 j_1^{} \sigma_1^{} \omega_{n_1^{}},j_2^{} \sigma_2^{}
 \omega_{n_2^{}}) \nonumber \\ & \quad  = 
 \delta_{n_1'\!+\!n_2'n_1^{~}\!+n_2^{~}} 
 \gamma_2(j_1' \sigma_1' ,j_2' \sigma_2';j_1^{} \sigma_1^{},j_2^{} \sigma_2^{}
 ;\Pi,\Chi,\Delta)
\end{flalign}

For convenience, we have here also listed the fermionic frequencies in terms of the bosonic ones:
\begin{subequations}
 \begin{align}
  \omega_{n_1'} = \tfrac{1}{2} (\Pi + \Chi + \Delta)\, , & &
  \omega_{n_2'} = \tfrac{1}{2} (\Pi - \Chi - \Delta) \, , \\
  \omega_{n_1^{~}} = \tfrac{1}{2} (\Pi + \Chi - \Delta)\, , & &
  \omega_{n_2^{~}} = \tfrac{1}{2} (\Pi - \Chi + \Delta)\, .
 \end{align}
\end{subequations}

\subsection{Coupled-Ladder Approximation}

The basic idea of the CLA scheme was introduced in
Refs.~\onlinecite{Karrasch2008,Jakobs2010} for the frequency
parametrization of the single-impurity Anderson model and was further
developed for inhomogeneous Fermi systems with onsite interaction in
Ref.~\onlinecite{Bauer2013}.  Here we will go one step further and
extend this scheme to treat interacting models with two-particle
interactions of finite range, using an idea similar to the singular
mode fRG approach introduced in \onlinecite{Husemann2009}. There, the
vertex structure in momentum space was decomposed into fermion
bilinears that interact via exchange bosons and it was shown that this
decomposition admits a systematic approximation by an expansion using
form factors. Here, we will proceed similar in position space,
introducing ``short indices'' $k,l$ that will control the extent of
our approximation and act similar to the mentioned form factor
expansion.

In the case of third-order truncated
fRG, BHD introduced two different approximation schemes. The simpler
``static second order fRG'' (sfRG2) neglects the frequency dependence
of the vertex; the more elaborate ``dynamic second order fRG'' (dfRG2)
includes the frequency dependence of the vertex within a channel
approximation, reducing this dependence from the generic
$\mathcal{O}(N_f^3)$ to $\mathcal{O}(N_f)$ where $N_f$ is the number
of used frequencies. In the case of the onsite model, it turned out
that static compared to dynamic fRG produces some artifacts concerning
the pinch-off point of the conductance of a QPC, but yields
essentially the same shape for the conductance steps as dynamic
fRG. For this reason and since it is a factor of $N_f$ cheaper, we
will only compute the static fRG flow in our numerical
work. Nevertheless, we will derive here the full dynamic flow
equations, and in principle, it should be no problem to implement
these too.

The dfRG2 scheme exploits the fact that the bare
vertex consists of a density-density interaction
\begin{equation}
\label{eq:bare-vertex}
\begin{array}{l}
v(j_1' \sigma_1',j_2' \sigma_2';j_1^{} \sigma_1^{},j_2^{} \sigma_2^{}) 
=\\ \;  \qquad \delta^{L_U}_{j_1^{}j_2^{}} U_{j_1^{}j_2^{}} 
\left[\left(1-\delta_{j_1^{}j_2^{}}\right)\delta_{\sigma_1^{}\sigma_2^{}} +
\delta_{\sigma_1^{}\bar{\sigma}_2^{}}
\right]\\
\; \qquad \times \left(\delta_{j_1'j_1^{}} \delta_{j_2'j_2^{}} \delta_{\sigma_1'\sigma_1^{}}
\delta_{\sigma_2'\sigma_2^{}}- 
\delta_{j_1'j_2^{}} \delta_{j_2'j_1^{}} \delta_{\sigma_1'\sigma_2^{}}
\delta_{\sigma_2'\sigma_1^{}} \right)\, , 
\end{array}
\end{equation}
and parametrizes the vertex in terms of
$\mathcal{O}(N^2L_U^2N_{\rm f})$ independent variables.
Here $\delta^{L_U}_{j_1j_2}$=1 if $|j_1-j_2|\leq L_U$ and is otherwise set to zero.

Using this vertex, we can now consider a simplified version of the vertex flow equation
\eqref{eqgamma2DGL}, where the feedback of the vertex flow is neglected: on the
r.h.s. we replace $\gamma_2^{\Lambda}\to v$. If the feedback of the self-energy
were also neglected, this would be equivalent to calculating the vertex in second order perturbation
theory. 
As a consequence, all generated vertex contributions have one of the following structures: 
\begin{subequations}
\label{eqStructures}
\begin{align}
\label{eqStructure_P}
P^{kl}_{ji\sigma \sigma'} (\Pi) &:= 
\gamma_p^\Lambda 
\left( \color{DGLorange} j\sigma , j\!+\!k\, \sigma' 
\color{black} ; 
\color{RoyalBlue} i\sigma , i\!+\!l\, \sigma' \color{black};\Pi \right)
\nonumber\\  
&\!\!\stackrel{\mathcal{O}(v^2)}{\simeq} 
\begin{matrix}
\begingroup%
  \makeatletter%
  \providecommand\color[2][]{%
    \errmessage{(Inkscape) Color is used for the text in Inkscape, but the package 'color.sty' is not loaded}%
    \renewcommand\color[2][]{}%
  }%
  \providecommand\transparent[1]{%
    \errmessage{(Inkscape) Transparency is used (non-zero) for the text in Inkscape, but the package 'transparent.sty' is not loaded}%
    \renewcommand\transparent[1]{}%
  }%
  \providecommand\rotatebox[2]{#2}%
  \ifx\svgwidth\undefined%
    \setlength{\unitlength}{135.45369873bp}%
    \ifx\svgscale\undefined%
      \relax%
    \else%
      \setlength{\unitlength}{\unitlength * \real{\svgscale}}%
    \fi%
  \else%
    \setlength{\unitlength}{\svgwidth}%
  \fi%
  \global\let\svgwidth\undefined%
  \global\let\svgscale\undefined%
  \makeatother%
  \begin{picture}(1,0.46845314)%
    \put(0,0){\includegraphics[width=\unitlength,page=1]{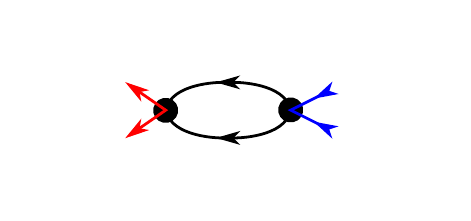}}%
    \put(0.02506257,0.37414264){\color[rgb]{0,0,0}\makebox(0,0)[lb]{\smash{$\Pi-\omega_{n'}$}}}%
    \put(0.1727145,0.28514053){\color[rgb]{0,0,0}\makebox(0,0)[lb]{\smash{$j\sigma$}}}%
    \put(0,0){\includegraphics[width=\unitlength,page=2]{pss.pdf}}%
    \put(0.02506257,0.1374886){\color[rgb]{0,0,0}\makebox(0,0)[lb]{\smash{$j+k\,\sigma'$}}}%
    \put(0.1727145,0.07883878){\color[rgb]{0,0,0}\makebox(0,0)[lb]{\smash{$\omega_{n'}$}}}%
    \put(0.3794272,0.37414264){\color[rgb]{0,0,0}\makebox(0,0)[lb]{\smash{$\Pi-\omega_{n''}$}}}%
    \put(0.46801836,0.31467092){\color[rgb]{0,0,0}\makebox(0,0)[lb]{\smash{$\sigma$}}}%
    \put(0.46801836,0.10795822){\color[rgb]{0,0,0}\makebox(0,0)[lb]{\smash{$\sigma'$}}}%
    \put(0.43848798,0.04930839){\color[rgb]{0,0,0}\makebox(0,0)[lb]{\smash{$\omega_{n''}$}}}%
    \put(0.73379183,0.28514053){\color[rgb]{0,0,0}\makebox(0,0)[lb]{\smash{$i\sigma$}}}%
    \put(0.73379183,0.1374886){\color[rgb]{0,0,0}\makebox(0,0)[lb]{\smash{$i+l\ \sigma'$ }}}%
    \put(0.73379183,0.07883878){\color[rgb]{0,0,0}\makebox(0,0)[lb]{\smash{$\omega_n$}}}%
    \put(0.73379183,0.37414264){\color[rgb]{0,0,0}\makebox(0,0)[lb]{\smash{$\Pi-\omega_n$}}}%
    \put(0,0){\includegraphics[width=\unitlength,page=3]{pss.pdf}}%
  \end{picture}%
\endgroup%

\end{matrix}\, , 
\\
\label{eqStructure_Pbar}
\bar{P}^{kl}_{ji\sigma \sigma'} (\Pi ) &:= 
\gamma_p^\Lambda \left( 
\color{DGLorange} j \sigma  , j\!+\!k\, \sigma' 
\color{black} ; 
\color{RoyalBlue} i \sigma'  , i\!+\!l \, \sigma  
\color{black},\Pi \right) 
\nonumber \\  
&\!\!\stackrel{\mathcal{O}(v^2)}{\simeq}
\begin{matrix}
\begingroup%
  \makeatletter%
  \providecommand\color[2][]{%
    \errmessage{(Inkscape) Color is used for the text in Inkscape, but the package 'color.sty' is not loaded}%
    \renewcommand\color[2][]{}%
  }%
  \providecommand\transparent[1]{%
    \errmessage{(Inkscape) Transparency is used (non-zero) for the text in Inkscape, but the package 'transparent.sty' is not loaded}%
    \renewcommand\transparent[1]{}%
  }%
  \providecommand\rotatebox[2]{#2}%
  \ifx\svgwidth\undefined%
    \setlength{\unitlength}{135.45369873bp}%
    \ifx\svgscale\undefined%
      \relax%
    \else%
      \setlength{\unitlength}{\unitlength * \real{\svgscale}}%
    \fi%
  \else%
    \setlength{\unitlength}{\svgwidth}%
  \fi%
  \global\let\svgwidth\undefined%
  \global\let\svgscale\undefined%
  \makeatother%
  \begin{picture}(1,0.46845314)%
    \put(0,0){\includegraphics[width=\unitlength,page=1]{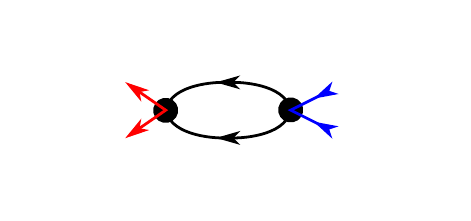}}%
    \put(0.02506257,0.37414264){\color[rgb]{0,0,0}\makebox(0,0)[lb]{\smash{$\Pi-\omega_{n'}$}}}%
    \put(0.1727145,0.28514053){\color[rgb]{0,0,0}\makebox(0,0)[lb]{\smash{$j\sigma$}}}%
    \put(0,0){\includegraphics[width=\unitlength,page=2]{barpss.pdf}}%
    \put(0.02506257,0.1374886){\color[rgb]{0,0,0}\makebox(0,0)[lb]{\smash{$j+k\,\sigma'$}}}%
    \put(0.1727145,0.07883878){\color[rgb]{0,0,0}\makebox(0,0)[lb]{\smash{$\omega_{n'}$}}}%
    \put(0.3794272,0.37414264){\color[rgb]{0,0,0}\makebox(0,0)[lb]{\smash{$\Pi-\omega_{n''}$}}}%
    \put(0.46801836,0.31467092){\color[rgb]{0,0,0}\makebox(0,0)[lb]{\smash{$\sigma$}}}%
    \put(0.46801836,0.10795822){\color[rgb]{0,0,0}\makebox(0,0)[lb]{\smash{$\sigma'$}}}%
    \put(0.43848798,0.04930839){\color[rgb]{0,0,0}\makebox(0,0)[lb]{\smash{$\omega_{n''}$}}}%
    \put(0.73379183,0.28514053){\color[rgb]{0,0,0}\makebox(0,0)[lb]{\smash{$i\sigma'$}}}%
    \put(0.73379183,0.1374886){\color[rgb]{0,0,0}\makebox(0,0)[lb]{\smash{$i+l\ \sigma$ }}}%
    \put(0.73379183,0.07883878){\color[rgb]{0,0,0}\makebox(0,0)[lb]{\smash{$\omega_n$}}}%
    \put(0.73379183,0.37414264){\color[rgb]{0,0,0}\makebox(0,0)[lb]{\smash{$\Pi-\omega_n$}}}%
    \put(0,0){\includegraphics[width=\unitlength,page=3]{barpss.pdf}}%
  \end{picture}%
\endgroup%

\end{matrix}\, , 
\\ 
\label{eqStructure_X}
X^{kl}_{ji\sigma \sigma'} ( \Chi ) &:= 
\gamma_x^\Lambda \left( 
\color{DGLorange} j\sigma , i\!+\!l\, \sigma' 
\color{black} ; \color{RoyalBlue} i\sigma , j\!+\!k\, \sigma'
\color{black};\Chi \right) 
\nonumber \\  
&\!\! \stackrel{\mathcal{O}(v^2)}{\simeq}
\begin{matrix}
\begingroup%
  \makeatletter%
  \providecommand\color[2][]{%
    \errmessage{(Inkscape) Color is used for the text in Inkscape, but the package 'color.sty' is not loaded}%
    \renewcommand\color[2][]{}%
  }%
  \providecommand\transparent[1]{%
    \errmessage{(Inkscape) Transparency is used (non-zero) for the text in Inkscape, but the package 'transparent.sty' is not loaded}%
    \renewcommand\transparent[1]{}%
  }%
  \providecommand\rotatebox[2]{#2}%
  \ifx\svgwidth\undefined%
    \setlength{\unitlength}{135.45369873bp}%
    \ifx\svgscale\undefined%
      \relax%
    \else%
      \setlength{\unitlength}{\unitlength * \real{\svgscale}}%
    \fi%
  \else%
    \setlength{\unitlength}{\svgwidth}%
  \fi%
  \global\let\svgwidth\undefined%
  \global\let\svgscale\undefined%
  \makeatother%
  \begin{picture}(1,0.46845314)%
    \put(0,0){\includegraphics[width=\unitlength,page=1]{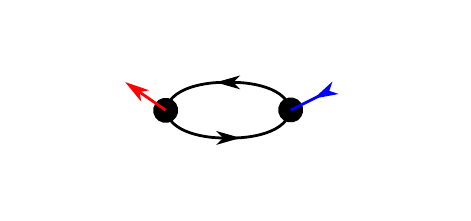}}%
    \put(0.02506257,0.37414264){\color[rgb]{0,0,0}\makebox(0,0)[lb]{\smash{$X+\omega_{n'}$}}}%
    \put(0.1727145,0.28514053){\color[rgb]{0,0,0}\makebox(0,0)[lb]{\smash{$j\sigma$}}}%
    \put(0,0){\includegraphics[width=\unitlength,page=2]{xsp.pdf}}%
    \put(0.02506257,0.1374886){\color[rgb]{0,0,0}\makebox(0,0)[lb]{\smash{$j+k\,\sigma'$}}}%
    \put(0.1727145,0.07883878){\color[rgb]{0,0,0}\makebox(0,0)[lb]{\smash{$\omega_{n'}$}}}%
    \put(0.3794272,0.37414264){\color[rgb]{0,0,0}\makebox(0,0)[lb]{\smash{$X+\omega_{n''}$}}}%
    \put(0.46801836,0.31467092){\color[rgb]{0,0,0}\makebox(0,0)[lb]{\smash{$\sigma$}}}%
    \put(0.46801836,0.10795822){\color[rgb]{0,0,0}\makebox(0,0)[lb]{\smash{$\sigma'$}}}%
    \put(0.43848798,0.04930839){\color[rgb]{0,0,0}\makebox(0,0)[lb]{\smash{$\omega_{n''}$}}}%
    \put(0.73379183,0.28514053){\color[rgb]{0,0,0}\makebox(0,0)[lb]{\smash{$i\sigma$}}}%
    \put(0.73379183,0.1374886){\color[rgb]{0,0,0}\makebox(0,0)[lb]{\smash{$i+l\,\sigma'$ }}}%
    \put(0.73379183,0.07883878){\color[rgb]{0,0,0}\makebox(0,0)[lb]{\smash{$\omega_n$}}}%
    \put(0.73379183,0.37414264){\color[rgb]{0,0,0}\makebox(0,0)[lb]{\smash{$X+\omega_n$}}}%
    \put(0,0){\includegraphics[width=\unitlength,page=3]{xsp.pdf}}%
    \put(-0.17919888,0.50000005){\color[rgb]{0,0,0}\makebox(0,0)[lt]{\begin{minipage}{0.14765193\unitlength}\raggedright \end{minipage}}}%
  \end{picture}%
\endgroup%

\end{matrix}\, , 
\\
\label{eqStructure_Xbar}
\bar{X}^{kl}_{ji\sigma \sigma'} ( \Chi ) &:=
\gamma_x^\Lambda \left( 
\color{DGLorange} j\sigma , i\!+\!l\, \sigma' 
\color{black} ; \color{RoyalBlue} i\sigma' , j\!+\!k\, \sigma
 \color{black};\Chi \right) 
\nonumber \\  
& \!\! \stackrel{\mathcal{O}(v^2)}{\simeq}
\begin{matrix}
\begingroup%
  \makeatletter%
  \providecommand\color[2][]{%
    \errmessage{(Inkscape) Color is used for the text in Inkscape, but the package 'color.sty' is not loaded}%
    \renewcommand\color[2][]{}%
  }%
  \providecommand\transparent[1]{%
    \errmessage{(Inkscape) Transparency is used (non-zero) for the text in Inkscape, but the package 'transparent.sty' is not loaded}%
    \renewcommand\transparent[1]{}%
  }%
  \providecommand\rotatebox[2]{#2}%
  \ifx\svgwidth\undefined%
    \setlength{\unitlength}{135.45369873bp}%
    \ifx\svgscale\undefined%
      \relax%
    \else%
      \setlength{\unitlength}{\unitlength * \real{\svgscale}}%
    \fi%
  \else%
    \setlength{\unitlength}{\svgwidth}%
  \fi%
  \global\let\svgwidth\undefined%
  \global\let\svgscale\undefined%
  \makeatother%
  \begin{picture}(1,0.46845314)%
    \put(0,0){\includegraphics[width=\unitlength,page=1]{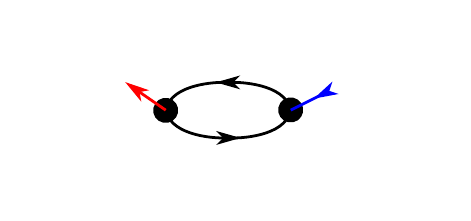}}%
    \put(0.02506257,0.37414264){\color[rgb]{0,0,0}\makebox(0,0)[lb]{\smash{$X+\omega_{n'}$}}}%
    \put(0.1727145,0.28514053){\color[rgb]{0,0,0}\makebox(0,0)[lb]{\smash{$j\sigma$}}}%
    \put(0,0){\includegraphics[width=\unitlength,page=2]{barxsp.pdf}}%
    \put(0.02506257,0.1374886){\color[rgb]{0,0,0}\makebox(0,0)[lb]{\smash{$j+k\,\sigma$}}}%
    \put(0.1727145,0.07883878){\color[rgb]{0,0,0}\makebox(0,0)[lb]{\smash{$\omega_{n'}$}}}%
    \put(0.3794272,0.37414264){\color[rgb]{0,0,0}\makebox(0,0)[lb]{\smash{$X+\omega_{n''}$}}}%
    \put(0.46801836,0.31467092){\color[rgb]{0,0,0}\makebox(0,0)[lb]{\smash{$\mu$}}}%
    \put(0.46801836,0.10795822){\color[rgb]{0,0,0}\makebox(0,0)[lb]{\smash{$\mu$}}}%
    \put(0.43848798,0.04930839){\color[rgb]{0,0,0}\makebox(0,0)[lb]{\smash{$\omega_{n''}$}}}%
    \put(0.73379183,0.28514053){\color[rgb]{0,0,0}\makebox(0,0)[lb]{\smash{$i\sigma'$}}}%
    \put(0.73379183,0.1374886){\color[rgb]{0,0,0}\makebox(0,0)[lb]{\smash{$i+l\,\sigma'$ }}}%
    \put(0.73379183,0.07883878){\color[rgb]{0,0,0}\makebox(0,0)[lb]{\smash{$\omega_n$}}}%
    \put(0.73379183,0.37414264){\color[rgb]{0,0,0}\makebox(0,0)[lb]{\smash{$X+\omega_n$}}}%
    \put(0,0){\includegraphics[width=\unitlength,page=3]{barxsp.pdf}}%
    \put(-0.17919888,0.50000005){\color[rgb]{0,0,0}\makebox(0,0)[lt]{\begin{minipage}{0.14765193\unitlength}\raggedright \end{minipage}}}%
  \end{picture}%
\endgroup%

\end{matrix}\, , 
\\
\label{eqStructure_D}
D^{kl}_{ji\sigma \sigma'} (\Delta ) &:=
\gamma_d^\Lambda 
\left(\color{DGLorange} j\sigma , i\!+\! l \, \sigma' 
\color{black} ; \color{RoyalBlue} j\!+\!k\,\sigma , i \sigma' 
\color{black} ; \Delta \right) 
\nonumber \\
&\!\!  \stackrel{\mathcal{O}(v^2)}{\simeq}
\begin{matrix}
\begingroup%
  \makeatletter%
  \providecommand\color[2][]{%
    \errmessage{(Inkscape) Color is used for the text in Inkscape, but the package 'color.sty' is not loaded}%
    \renewcommand\color[2][]{}%
  }%
  \providecommand\transparent[1]{%
    \errmessage{(Inkscape) Transparency is used (non-zero) for the text in Inkscape, but the package 'transparent.sty' is not loaded}%
    \renewcommand\transparent[1]{}%
  }%
  \providecommand\rotatebox[2]{#2}%
  \ifx\svgwidth\undefined%
    \setlength{\unitlength}{138.49735718bp}%
    \ifx\svgscale\undefined%
      \relax%
    \else%
      \setlength{\unitlength}{\unitlength * \real{\svgscale}}%
    \fi%
  \else%
    \setlength{\unitlength}{\svgwidth}%
  \fi%
  \global\let\svgwidth\undefined%
  \global\let\svgscale\undefined%
  \makeatother%
  \begin{picture}(1,0.51568129)%
    \put(0,0){\includegraphics[width=\unitlength,page=1]{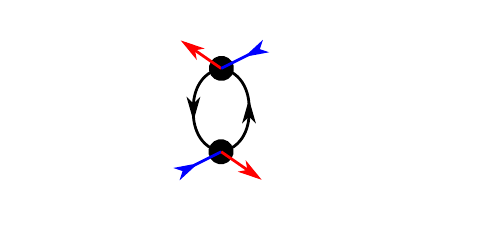}}%
    \put(0.28432468,0.42316151){\color[rgb]{0,0,0}\makebox(0,0)[lb]{\smash{$j\sigma$}}}%
    \put(0.57313885,0.42316151){\color[rgb]{0,0,0}\makebox(0,0)[lb]{\smash{$j+k\,\sigma$}}}%
    \put(0.25544326,0.10546591){\color[rgb]{0,0,0}\makebox(0,0)[lb]{\smash{$i\sigma'$}}}%
    \put(0.57313885,0.10546591){\color[rgb]{0,0,0}\makebox(0,0)[lb]{\smash{$i+l\,\sigma'$ }}}%
    \put(0.5749792,0.03372775){\color[rgb]{0,0,0}\makebox(0,0)[lb]{\smash{$\omega_n$}}}%
    \put(0.83395764,0.42356349){\color[rgb]{0,0,0}\makebox(0,0)[lb]{\smash{$\omega_{n'}$}}}%
    \put(0.13950157,0.03360083){\color[rgb]{0,0,0}\makebox(0,0)[lb]{\smash{$\omega_n+\Delta$}}}%
    \put(0.02439192,0.42356342){\color[rgb]{0,0,0}\makebox(0,0)[lb]{\smash{$\Delta+\omega_{n'}$}}}%
    \put(0.57313885,0.27875442){\color[rgb]{0,0,0}\makebox(0,0)[lb]{\smash{$\mu$}}}%
    \put(0.65978311,0.27915633){\color[rgb]{0,0,0}\makebox(0,0)[lb]{\smash{$\Delta+\omega_{n''}$}}}%
    \put(0.28432468,0.27875442){\color[rgb]{0,0,0}\makebox(0,0)[lb]{\smash{$\mu$}}}%
    \put(0.13991759,0.27915633){\color[rgb]{0,0,0}\makebox(0,0)[lb]{\smash{$\omega_{n''}$}}}%
    \put(0,0){\includegraphics[width=\unitlength,page=2]{dsp.pdf}}%
  \end{picture}%
\endgroup%

\end{matrix}\, , 
\\
\label{eqStructure_Dbar}
\bar{D}^{kl}_{ji\sigma \sigma'} (\Delta ) &:=
\gamma_d^\Lambda 
\left(\color{DGLorange} j\sigma , i\!+\! l \, \sigma' 
\color{black} ; \color{RoyalBlue} j\!+\!k\,\sigma' , i \sigma 
  \color{black} ; \Delta \right) 
\nonumber \\ 
&\!\! \stackrel{\mathcal{O}(v^2)}{\simeq}
\begin{matrix}
\begingroup%
  \makeatletter%
  \providecommand\color[2][]{%
    \errmessage{(Inkscape) Color is used for the text in Inkscape, but the package 'color.sty' is not loaded}%
    \renewcommand\color[2][]{}%
  }%
  \providecommand\transparent[1]{%
    \errmessage{(Inkscape) Transparency is used (non-zero) for the text in Inkscape, but the package 'transparent.sty' is not loaded}%
    \renewcommand\transparent[1]{}%
  }%
  \providecommand\rotatebox[2]{#2}%
  \ifx\svgwidth\undefined%
    \setlength{\unitlength}{138.49735718bp}%
    \ifx\svgscale\undefined%
      \relax%
    \else%
      \setlength{\unitlength}{\unitlength * \real{\svgscale}}%
    \fi%
  \else%
    \setlength{\unitlength}{\svgwidth}%
  \fi%
  \global\let\svgwidth\undefined%
  \global\let\svgscale\undefined%
  \makeatother%
  \begin{picture}(1,0.51568124)%
    \put(0,0){\includegraphics[width=\unitlength,page=1]{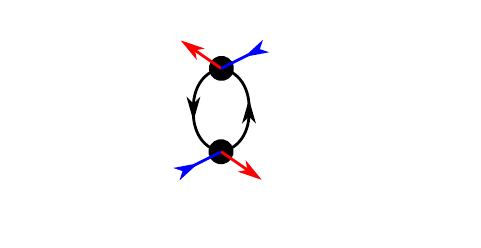}}%
    \put(0.28432468,0.42316129){\color[rgb]{0,0,0}\makebox(0,0)[lb]{\smash{$j\sigma$}}}%
    \put(0.57313885,0.42316129){\color[rgb]{0,0,0}\makebox(0,0)[lb]{\smash{$j+k\,\sigma'$}}}%
    \put(0.25544326,0.10546569){\color[rgb]{0,0,0}\makebox(0,0)[lb]{\smash{$i\sigma$}}}%
    \put(0.57313885,0.10546569){\color[rgb]{0,0,0}\makebox(0,0)[lb]{\smash{$i+l\,\sigma'$ }}}%
    \put(0.57313885,0.0336006){\color[rgb]{0,0,0}\makebox(0,0)[lb]{\smash{$\omega_n$}}}%
    \put(0.83406421,0.42356341){\color[rgb]{0,0,0}\makebox(0,0)[lb]{\smash{$\omega_{n'}$}}}%
    \put(0.14175789,0.0336006){\color[rgb]{0,0,0}\makebox(0,0)[lb]{\smash{$\omega_n+\Delta$}}}%
    \put(0.02439192,0.4235632){\color[rgb]{0,0,0}\makebox(0,0)[lb]{\smash{$\Delta+\omega_{n'}$}}}%
    \put(0.57313885,0.2787542){\color[rgb]{0,0,0}\makebox(0,0)[lb]{\smash{$\sigma$}}}%
    \put(0.65978311,0.27915611){\color[rgb]{0,0,0}\makebox(0,0)[lb]{\smash{$\Delta+\omega_{n''}$}}}%
    \put(0.28432468,0.2787542){\color[rgb]{0,0,0}\makebox(0,0)[lb]{\smash{$\sigma'$}}}%
    \put(0.13991759,0.27915611){\color[rgb]{0,0,0}\makebox(0,0)[lb]{\smash{$\omega_{n''}$}}}%
    \put(0,0){\includegraphics[width=\unitlength,page=2]{bardsp.pdf}}%
  \end{picture}%
\endgroup%

\end{matrix} , 
\end{align}
\end{subequations}

These terms depend only on a single bosonic frequency. The upper indices $^{kl}$,
are taken to run over the range
\begin{equation}
 -L \leq k,l \leq L \, , 
\end{equation}
where the control parameter $L$ sets the ``spatial feedback range''.
The bounds on the lower indices depend on the upper indices: if one of the site
indices of $\gamma_2$ lies outside the region $[-N',N']$ where $N'$ is defined by $N=2N'+1$, $\gamma_2$ is zero. 
Therefore, $i,j$ run between
\begin{align}
 \max(-N',-N'-l) \leq &\,i \leq \min(N',N'-l), \\
 \max(-N',-N'-k) \leq &\,j \leq \min(N',N'-k).
\end{align}

Analogously to BHD, we now feed back all those terms on the r.h.s.\ of
the flow equation \eqref{eqgamma2DGL} which conserve the site and spin structure indicated in Eq.~\eqref{eqStructures}.  As a first consequence, 
each vertex quantity is fully fed back into its own flow equation. Secondly, the
feedback between different quantities is restricted to those site indices which
have the appropriate structure. Furthermore, to avoid frequency mixing, the feedback
to a given channel from the other two channels is restricted to using only the static, i.e. zero frequency
component of the latter.

This scheme can be expressed by the replacement 
\begin{equation}
\gamma_2 \to \tilde{\gamma}_a
\label{eqApprScheme}
\end{equation}
on the r.h.s.\ of channel $a=p,x,d$ in Eq.~\eqref{eqgamma2DGL} where
$\tilde{\gamma}_a$ is defined as:
\begin{subequations}
\label{eqFeedback}
\begin{align}
& \tilde{\gamma}_p (j_1' \sigma_1', j_2' \sigma_2';
 	j_1^{} \sigma_1^{}, j_2^{} \sigma_2^{},\Pi) \nonumber \\
&\quad=\delta^L_{j_1'j_2'}\delta^L_{j_1^{}j_2^{}}\gamma_2 (j_1' \sigma_1', j_2' \sigma_2';
j_1^{} \sigma_1^{}, j_2^{} \sigma_2^{};\Pi,0,0) 
\\
& \tilde{\gamma}_x (j_1' \sigma_1', j_2' \sigma_2';
 	j_1^{} \sigma_1^{}, j_2^{} \sigma_2^{},\Chi) \nonumber \\
&\quad=\delta^L_{j_1'j_2^{}}\delta^L_{j_2'j_1^{}}\gamma_2 (j_1' \sigma_1', j_2' \sigma_2';
j_1^{} \sigma_1^{}, j_2^{} \sigma_2^{};0,\Chi,0) 
\\
& \tilde{\gamma}_d (j_1' \sigma_1', j_2' \sigma_2';
 	j_1^{} \sigma_1^{}, j_2^{} \sigma_2^{},\Delta) \nonumber \\
&\quad=\delta^L_{j_1'j_1^{}}\delta^L_{j_2'j_2^{}}\gamma_2 (j_1' \sigma_1', j_2' \sigma_2';
j_1^{} \sigma_1^{}, j_2^{} \sigma_2^{};0,0,\Delta) 
\end{align}
\end{subequations}

\subsection{Symmetries}
As can readily be checked, these flow equations respect the following symmetry
relations: 
\vspace{-2mm}
\begin{subequations}
\label{eq:vertexsymmetries}
\begin{align}
\label{eq:G-symmetry}
\mathcal{G}^{\sigma \Lambda}_{ij} (\omega_n) &= \mathcal{G}^{\sigma \Lambda}_{ji}
(\omega_n)
= \left[ \mathcal{G}^{\sigma \Lambda}_{ij} (-\omega_n) \right]^* \; , 
\\
\Sigma^{\sigma \Lambda}_{ij} (\omega_n)  &= \Sigma^{\sigma \Lambda}_{ji} (\omega_n)
= \left[ \Sigma^{\sigma \Lambda}_{ij} (- \omega_n) \right]^* \; , 
\end{align}
\end{subequations}
\begin{subequations}
\begin{align}
P^{kl}_{j i\sigma \sigma'}(\Pi) & = P^{lk}_{i j\sigma \sigma'}(\Pi) = 
P^{(\!-\!k)(\!-\!l)}_{(j\!+\!k) (i\!+\!l)\sigma'\sigma}(\Pi)\; ,  \nonumber\\
\bar{P}^{kl}_{j i\sigma \sigma'}(\Pi) & = \bar{P}^{lk}_{i j\sigma' \sigma}(\Pi) = 
\bar{P}^{(\!-\!k) (\!-\!l)}_{(j\!+\!k) (i\!+\!l)\sigma' \sigma}(\Pi)   \; , \nonumber\\
P^{kl}_{j i\sigma \sigma'}(\Pi)\!  &  = -\bar{P}^{-\!k l}_{j\!+\!k i\sigma' \sigma}(\Pi)
= -\bar{P}^{k (\!-\!l)}_{j (i\!+\!l)\sigma \sigma'}(\Pi)   \; , \nonumber \\
P^{}_{\sigma \sigma}   &  = \bar{P}^{}_{\sigma \sigma} \; ,
\end{align}
\begin{align} 
X^{kl }_{j i\sigma \sigma'}(X)   &=  X^{lk}_{i j\sigma \sigma'}(X) =
\Bigl[X^{(\!-\!k)(\!-\!l)}_{(j\!+\!k) (i\!+\!l)\sigma'\sigma}(X) \Bigr]^* \, , \nonumber\\ 
\bar{X}^{kl }_{j i\sigma \sigma'}(X)   &=  \bar{X}^{lk}_{i j\sigma' \sigma}(X) =
[\bar{X}^{(\!-\!k)(\!-\!l)}_{(j\!+\!k) (i\!+\!l)\sigma\sigma'}(X)]^* \, , \nonumber\\ 
X^{}_{\sigma \sigma}&= \bar{X}^{}_{\sigma\sigma} \; , 
\end{align}
\begin{align}
X  &= - \bar{D}\, ,  \quad \bar{X} = -D\, ,
\end{align}
\begin{align}
P^{kl}_{j i\sigma \sigma'} (\Pi )  & = \left[ P^{kl}_{j i\sigma \sigma'}
(-\Pi) \right]^* ,\nonumber\\ 
X^{kl}_{j i\sigma \sigma'}  (\Chi)    & =  
\big[ X^{kl}_{j i\sigma \sigma'} (-\Chi) \big]^*  \, , \nonumber\\
\bar{X}^{kl}_{j i\sigma \sigma'}  (\Delta)   &  =  
\big[ \bar{X}^{kl}_{j i\sigma \sigma'} (-\Delta) \big]^*  \, . 
\end{align}
\end{subequations}
As a result, all relevant information is contained in a small number of independent frequency-dependent block-matrices,
which we define as follows:
\begin{align}
\label{eqNonzeroVertex}
P_{}^\Lambda &= P_{\uparrow \downarrow}^\Lambda \; , \nonumber  
P_{\sigma }^{\Lambda}  = P_{\sigma \sigma}^\Lambda \; , \nonumber \\ 
X_{}^\Lambda  &= X_{\uparrow \downarrow}^\Lambda \; ,\\ 
D_{}^\Lambda &= D_{\uparrow \downarrow}^\Lambda \;, 
D^{\Lambda}_{\sigma }  = D_{\sigma \sigma}^\Lambda  \nonumber \; ,
\end{align}
where the superscript $\Lambda$ signifies a dependence on the flow parameter.

The flow equations for these matrices can be derived starting from
Eqs.~\eqref{eqgamma2DGL}. The replacement
\eqref{eqApprScheme} restricts the internal quantum numbers on the r.h.s.\  of the flow
equation $q_3^{}$, $q_4^{}$, $q_3'$, and $q_4'$ according to
the definitions \eqref{eqStructures}: 
\begin{widetext}
\begin{subequations}
\label{eq:derive-flow-equations-channels-explicit}  
\begin{align}
\label{eq:derive-flow-equations-P-channel}
\dot P^{kl\Lambda}_{ji} (\Pi) & = 
\dot{\gamma}_p^\Lambda \left( \color{DGLorange} 
j\uparrow , j\!+\!k \downarrow 
\color{black} ; \color{RoyalBlue} 
i\uparrow , i\!+\!l \downarrow
\color{black};\Pi \right) \\ \nonumber
& = T \sum_{j'i'k'l',n} 
\Bigl[ 
 \tilde{\gamma}_p^\Lambda \left( \color{DGLorange}
j\uparrow , j\!+\!k \, \downarrow 
\color{black};
i'\uparrow , i'\!+\!l' \downarrow; \Pi
\right)
\mathcal{S}^{\uparrow \Lambda}_{i'j'}(\omega_{n}) 
\mathcal{G}^{\downarrow \Lambda}_{i'\!+\!l' j'\!+\!k'}(\Pi\!-\!  \omega_{n})
\tilde{\gamma}_p^\Lambda \left( 
j'\uparrow , j'\!+\!k' \downarrow ;
\color{RoyalBlue} 
i\uparrow , i\!+\!l \downarrow 
\color{black} ; \Pi
\right)
\Bigr.
\\ \nonumber 
&\qquad \qquad \quad \Bigl.+ 
 \tilde{\gamma}_p^\Lambda \left( \color{DGLorange}
j\uparrow , j\!+\!k \, \downarrow 
\color{black};
i'\downarrow , i'\!+\!l' \uparrow; \Pi
\right)
\mathcal{S}^{\downarrow \Lambda}_{i' j'}(\omega_{n}) 
\mathcal{G}^{\uparrow \Lambda}_{i'\!+\!l' j'\!+\!k'}(\Pi\!-\!  \omega_{n})
\tilde{\gamma}_p^\Lambda \left( 
j'\downarrow , j'\!+\!k' \uparrow ;
\color{RoyalBlue} 
i\uparrow , i\!+\!l \downarrow 
\color{black} ; \Pi
\right)
\Bigr] \, ,
\\  
\label{eq:derive-flow-equations-Ps-channel}
\dot P^{kl\Lambda}_{ji\sigma} (\Pi) & = 
\dot{\gamma}_p^\Lambda \left( \color{DGLorange} 
j\sigma , j\!+\!k \sigma 
\color{black} ; \color{RoyalBlue} 
i\sigma , i\!+\!l \sigma
\color{black};\Pi \right) \\ \nonumber
& = T \sum_{j'i'k'l',n} 
 \tilde{\gamma}_p^\Lambda \left( \color{DGLorange}
j\sigma , j\!+\!k \, \sigma 
\color{black};
i'\sigma , i'\!+\!l' \sigma; \Pi
\right)
\mathcal{S}^{\sigma \Lambda}_{i'j'}(\omega_{n}) 
\mathcal{G}^{\sigma \Lambda}_{i'\!+\!l' j'\!+\!k'}(\Pi\!-\!  \omega_{n})
\tilde{\gamma}_p^\Lambda \left( 
j'\sigma , j'\!+\!k' \sigma ;
\color{RoyalBlue} 
i\sigma , i\!+\!l \sigma 
\color{black} ; \Pi
\right)
\\  
\label{eq:derive-flow-equations-X-channel}
\dot X^{kl\Lambda}_{ji} (\Chi) &= 
\dot{\gamma}_x^\Lambda \left( \color{DGLorange} 
j\uparrow , i\!+\!l \, \downarrow 
\color{black} ; \color{RoyalBlue} 
i\uparrow  , j\!+\!k\, \downarrow 
\color{black}; \Chi
\right) \\ \nonumber
&=  T \sum_{i'j'l'k',n}
 \tilde{\gamma}_x^\Lambda \left( \color{DGLorange}
 j\uparrow 
\color{black} ,  
i'\!+\!l'\, \downarrow ;  
i' \uparrow  , 
\color{RoyalBlue}
 j\!+\!k\, \downarrow \color{black} ; \Chi \right)
\left[\mathcal{S}^{\uparrow \Lambda}_{i' j'}(\omega_{n}\!+\Chi) 
\mathcal{G}^{\downarrow \Lambda}_{j'\!+k' i'\!+l'}(\omega_{n}) 
+\mathcal{S}^{\downarrow \Lambda}_{j'\!+\!k'i'\!+\!l'}( \omega_{n}) 
\mathcal{G}^{\uparrow \Lambda}_{i'j'}(\omega_{n}\!+ \Chi) \right] \\
&\quad \quad \qquad \ \tilde{\gamma}_x^\Lambda \left( j'\uparrow ,  \color{DGLorange} 
i\!+\!l\,\downarrow \nonumber 
\color{black} ;  
\color{RoyalBlue} i\uparrow  \color{black}, 
j'\!+\!k'\, \downarrow  ;\Chi \right)\Bigr.
\\ \nonumber
\label{eq:derive-flow-equations-D-channel}
\dot{D}^{kl\Lambda}_{ji\sigma \sigma'} (\Chi) &= 
\dot{\gamma}_d^\Lambda \left( \color{DGLorange} 
j\sigma , i\!+\!l \, \sigma' 
\color{black} ; \color{RoyalBlue} 
j\!+\!k\, \sigma , i \sigma'
\color{black}; \Delta
\right) \\ \nonumber
&= - T \sum_{\begin{smallmatrix}i'j'l'k'\\n,\sigma''\end{smallmatrix}}
 \tilde{\gamma}_d^\Lambda \left( \color{DGLorange}
 j\sigma 
\color{black} ,  
i'\!+\!l'\, \sigma'' ;  
\color{RoyalBlue}
 j\!+\!k\, \sigma \color{black} 
,i' \sigma''   
 ; \Delta \right)
\left[\mathcal{S}^{\sigma'' \Lambda}_{i'\!+\!l'j'\!+\!k'}(\omega_{n}) 
\mathcal{G}^{\sigma'' \Lambda}_{i'j'}(\omega_{n}\!+\!\Delta ) 
+\mathcal{G}^{\sigma'' \Lambda}_{i'\!+\!l'j'\!+\!k'}(\omega_{n}) 
\mathcal{S}^{\sigma'' \Lambda}_{i'j'}(\omega_{n}\!+\!\Delta)\right] \\ \nonumber 
&\quad \quad \qquad \quad \tilde{\gamma}_d^\Lambda \left( j'\sigma'' ,  \color{DGLorange} 
i\!+\!l\,\sigma' 
\color{black} ;  
j'\!+\!k'\, \sigma'' ,
\color{RoyalBlue} i\sigma'  \color{black} 
 ;\Delta \right)
\\ 
\end{align}
\end{subequations}
\end{widetext}
The initial conditions are
\begin{align}
&P^{\Lambda_i}=P^{\Lambda_i}_\sigma =X^{\Lambda_i} =
D^{\Lambda_i}_{\sigma \sigma'} = 0 \; . 
\end{align}

These equations can be compactly
written in block-matrix form
\begin{subequations}
\label{eq:matrixODE}
\begin{align}
\label{eq:PODE}
\frac{d}{d\Lambda} P^{\Lambda} (\Pi)  = &
\tilde{P}^{\Lambda} (\Pi) \cdot \bubble^{p \Lambda} (\Pi) \cdot \tilde{P}^{\Lambda} (\Pi) \; ,\\
\label{eq:PsODE}
\frac{d}{d\Lambda} P^{\Lambda}_\sigma (\Pi)  = &
\tilde{P}^{\Lambda}_\sigma (\Pi) \cdot \bubble^{p \Lambda}_\sigma (\Pi) \cdot
\tilde{P}^{\Lambda}_\sigma (\Pi) \; ,\\
\label{eq:XODE}
\frac{d}{d\Lambda} X^{\Lambda} (\Chi)  = &
\tilde{X}^{\Lambda} (\Chi) \cdot \bubble^{x \Lambda} (\Chi) \cdot \tilde{X}^{\Lambda} (\Chi) \; ,\\
\label{eq:DODE}
\frac{d}{d\Lambda} D^{\Lambda}_{\sigma \sigma'} (\Delta)  = &- \sum_{\sigma''}
\tilde{D}^{\Lambda}_{\sigma \sigma''} (\Delta) \cdot 
\bubble^{d \Lambda}_{\sigma''} (\Delta) \cdot 
\tilde{D}^{\Lambda}_{\sigma'' \sigma'}
(\Delta) \, ,
\end{align}
\end{subequations}
where '$\cdot$' denotes a block-matrix multiplication:
\begin{equation}
 \left[A\cdot B\right]_{ji}^{kl} = \sum_{j'k'} A_{j j'}^{kk'} B_{j'i}^{k'l} \, 
\end{equation}
and we have introduced the definitions 
\begin{widetext}
\begin{subequations}
\label{eq:crossfeed}
\begin{align}
\label{eq:crossfeed_P}
\tilde{P}^{kl\Lambda}_{ji} (\Pi) = \,&
\tilde{\gamma}_p^\Lambda \left(  
j\uparrow , j\!+\!k \downarrow ; 
i\uparrow , i\!+\!l \downarrow
;\Pi \right) \nonumber  \\
= \, & \delta_{ji} \delta_{kl} U_{j  j+k} 
+ P^{kl\Lambda}_{ji}  (\Pi) 
 + \delta_{j i+l}^L \delta_{i j+k}^L
X^{(i+l-j)(j+k-i)\Lambda}_{ji}  (0) 
 + \delta_{i j}^L \delta_{j+k i+l}^L
D^{(i-j)(j+k-i-l)\Lambda}_{j(i+l) \uparrow \downarrow}  (0)
\, ,
\\
\label{eq:crossfeed_Ps}
\tilde{P}^{kl\Lambda}_{ji\sigma} (\Pi) = \,&
\tilde{\gamma}_p^\Lambda \left(  
j\sigma , j+k \sigma ; 
i\sigma , i+l \sigma
;\Pi \right) \nonumber \\
= \, & \delta_{ji} \delta_{kl} U_{jj+k} 
- \delta_{k,-l} \delta_{(j+k)i} U_{ji}
+ P^{kl\Lambda}_{ji\sigma}  (\Pi) 
 - \delta_{i+l j}^L \delta_{j+k i}^L
D^{(i+l-j)(j+k-i)\Lambda}_{ji\sigma}  (0) 
 + \delta_{i j}^L \delta_{j+k i+l}^L
D^{(i-j)(j+k-i-l)\Lambda}_{j(i+l) \sigma}  (0)
\, ,
\\
\label{eq:crossfeed_X}
\tilde{X}^{kl\Lambda}_{ji} (\Chi) = \,&
\tilde{\gamma}_x^\Lambda \left(  
j\uparrow , i+l \downarrow ; 
i\uparrow , j+k \downarrow
;\Chi \right) \nonumber \\
= \, & \delta_{ji} \delta_{kl} U_{jj+k} 
+ X^{kl\Lambda}_{ji}  (\Chi) 
 + \delta_{i+l j}^L \delta_{j+k i}^L
P^{(i+l-j)(j+k-i)\Lambda}_{ji}  (0) 
 + \delta_{i j}^L \delta_{j+k i+l}^L
D^{(i-j)(i+l-j-k)\Lambda}_{j(j+k) \uparrow \downarrow}  (0)
\, ,
\\
\label{eq:crossfeed_D}
\tilde{D}^{kl\Lambda}_{ji \sigma \sigma'} (\Delta) = \,&
\tilde{\gamma}_d^\Lambda \left(  
j\sigma , i+l \sigma' ; 
j+k \sigma , i \sigma'
;\Delta \right) \nonumber \\
= \, & \delta_{0k} \delta_{0l} U_{ji}
 - \delta_{\sigma \sigma'}\delta_{ji} \delta_{kl} U_{jj+k} 
+ D^{kl\Lambda}_{ji \sigma \sigma'}  (\Delta) 
 + \delta_{i+l j}^L \delta_{j+k i}^L
P^{(i+l-j)(i-j-k)\Lambda}_{j(j+k) \sigma \sigma'}  (0) 
 + \delta_{i j}^L \delta_{j+k i+l}^L
X^{(i-j)(i+l-j-k)\Lambda}_{j(j+k) \sigma \sigma'}  (0)
\, ,
\end{align}
\end{subequations}
\end{widetext}
which account for the inter-channel feedback contained in equation~\eqref{eqApprScheme}.
Note that Eq.~\eqref{eq:crossfeed_D} is not fully expressed in terms
of the definitions~\eqref{eqNonzeroVertex}. This can only been done once $\sigma$ and $\sigma'$ are
specified explicitly and then leads to three independent equations.
$\bubble^p$, $\bubble^x$ and $\bubble^d$ each represent a specific bubble,
i.e.\ a product of two propagators summed over an internal frequency:
\begin{subequations}
\label{eq:bubble}
\begin{align}
\label{eq:Pip}
\bubble^{lk, p \Lambda}_{ij} (\Pi) &=T \sum_{n} \Bigl[ 
\mathcal{S}^{\uparrow \Lambda}_{ij}(\omega_{n}) 
\mathcal{G}^{\downarrow \Lambda}_{i\!+\!l j\!+\!k}(\Pi\!-\!  \omega_{n})
\Bigr.\nonumber \\
& \qquad \qquad \Bigl.
+ \mathcal{S}^{\downarrow \Lambda}_{i\!+\!l j\!+\!k}(\omega_{n}) 
\mathcal{G}^{\uparrow \Lambda}_{i j}(\Pi\!-\!  \omega_{n})
\Bigr]\\
\label{eq:Pip2}
\bubble^{lk, p \Lambda}_{ij \sigma} (\Pi) &=T \sum_{n} \Bigl[ 
\mathcal{S}^{\sigma \Lambda}_{ij}(\omega_{n}) 
\mathcal{G}^{\sigma \Lambda}_{i\!+\!l j\!+\!k}(\Pi\!-\!  \omega_{n})
\Bigr]\\
\label{eq:Pix}
\bubble^{lk,x \Lambda}_{ij} (\Chi )&=T\sum_{n} 
\Bigl[ 
\mathcal{S}^{\downarrow \Lambda}_{i\!+\!lj\!+\!k}(\omega_{n}) 
\mathcal{G}^{\uparrow \Lambda}_{ij}(\omega_{n}\!+\!\Chi ) 
\nonumber \\
&
\phantom{=T\sum_{n^{}}\Bigl[} + 
\mathcal{G}^{\downarrow \Lambda}_{j\!+\!ki\!+\!l}( \omega_{n}) 
\mathcal{S}^{\uparrow \Lambda}_{ij}(\omega_{n} \!+\! \Chi ) 
\Bigr],\\
\label{eq:Pid}
\bubble^{lk,d \Lambda}_{ij \sigma} (\Delta )&=T\sum_{n} 
\Bigl[ 
\mathcal{S}^{\sigma \Lambda}_{i\!+\!lj\!+\!k}(\omega_{n}) 
\mathcal{G}^{\sigma \Lambda}_{ij}(\omega_{n}\!+\!\Delta ) 
\nonumber \\
&
\phantom{=T\sum_{n^{}}\Bigl[} + 
\mathcal{G}^{\sigma \Lambda}_{i\!+\!lj\!+\!k}( \omega_{n} ) 
\mathcal{S}^{\sigma \Lambda}_{ij}(\omega_{n} \!+\! \Delta ) 
\Bigr].
\end{align}
\end{subequations}
\subsection{eCLA vs. CLA and the Role of $D_{\uparrow \downarrow}$}
\begin{figure}
   \includegraphics[width=83mm]{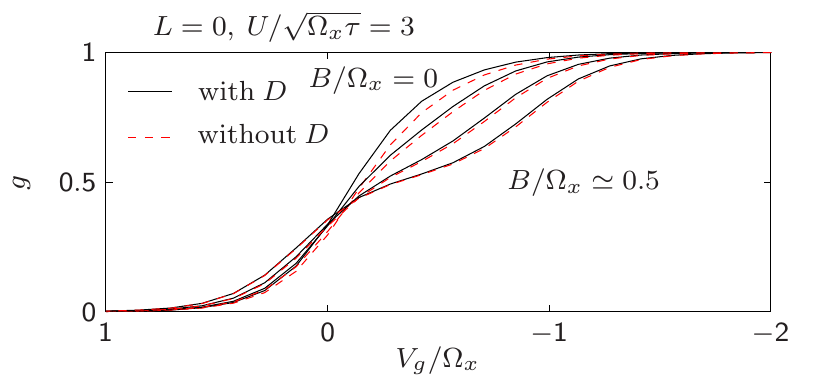} 
   \caption{\small The linear conductance $g = G/G_Q$ of a QPC as
     function of gate voltage, plotted for the cases with and without
     feedback of $D^{\uparrow\downarrow}$ in an intermediate parameter
     regime for four equidistant magnetic fields. Note that the
     difference between the two cases is suppressed with increasing
     the magnetic field.  }
\label{Paper_ohne_D2}
\end{figure}

\changed{Let us now recapitulate the
  similarities and differences between our new eCLA method to the
  previous CLA method used in Ref.~\onlinecite{Bauer2014}. There, only
  onsite models were considered and the guiding idea for
  approximations in the fRG flow was to include only those vertex
  structures which are already generated in second order in the
  interaction. Therefore it was sufficient to consider only an onsite
  feedback between the individual channels, i.e. the feedback range
  was the same as the interaction range. In the development of the
  eCLA we followed the same idea, but found it to be advantagous to
  separate the feedback length $L$ from the actual range of the
  interaction $L_U$. To be exact in second order, $L$ has to be chosen
  at least as large as $L_U$. However, it can be chosen also larger
  than $L_U$, and thus enables us to study the importance of the
  neglected higher order terms. If $L$ is chosen exactly equal to
  $L_U$, we are in principle back at the original idea to include only
  vertex structures in the flow which are already generated in second
  order of the interaction. 

  \changed{However, there is one exception to the last statement: for
    purely onsite interactions ($L_U=0$), the contributions of
    $D^{\uparrow \downarrow}$ and $P^{\sigma \sigma}$ to the vertex
    are of third and fourth order, respectively. In
    Ref.~\onlinecite{Bauer2014} they were therefore neglected,
    consistent with the policy of keeping only structures generated in
    second order. In the present paper, however, our implementation
    does not explicitly distinguish between $L_U=0$ and $L_U>0$ and
    includes the $D^{\uparrow \downarrow}$ and $P^{\sigma \sigma}$
    contributions regardless of the values of $L_U$ and $L$, even
    for $L_U=L=0$. To be specific, for $L_U=L=0$ our present
    flow scheme keeps $P^{\sigma \sigma}=0$ but leads to a finite
    contribution of $D^{\uparrow \downarrow}$.  Consequently, our
    results for $L_U=L=0$ differ slightly from those obtained in
    Refs.~\onlinecite{Bauer2013,Bauer2014}, and the difference is a
    measure of the magnitude of the third-order
    $D^{\uparrow\downarrow}$ contribution.}  

In
    Fig.~\ref{Paper_ohne_D2}, we compare the dependence of the QPC
    conductance on the magnetic field for a model with purely onsite
    interactions (defined in Sec.~\ref{secfRGresults} below) for both
    CLA and eCLA with $L=0$. The difference is most noticeable for
    $B=0$ in the region of the 0.7 anomly, i.e. in the regime where
    interactions influence the conductance most strongly, but even here the difference is not very
    big. (Of course this holds only in intermediate parameter regimes,
    i.e.\ in regimes where both the eCLA and the CLA are convergent.)}

\subsection{The flow equation of the self-energy}

Using the above definitions, the flow equation of the self-energy,
\eqref{eqgamma1DGL}, can be written explicitly as 
\begin{align} \label{eqSelfExplicit} \frac{d}{d\Lambda}
\Sigma_{ji}^{\sigma \Lambda} (\omega_n) = - T & \sum_{k,\sigma',n'} \Bigl\{ 
\sum_l \mathcal{S}_{i+l,j+k}^{\sigma' \Lambda}(\omega_n')\Bigl[U_{i(i+l)}\delta_{lk}\delta_{ji} \nonumber \\
&-U_{ij}\delta_{k,-l}\delta_{j(i+l)}\delta_{\sigma\sigma'}+P^{kl\Lambda}_{ji\sigma\sigma'}(\omega_n+\omega_n') \nonumber \\
&+X^{ kl\Lambda}_{ji\sigma\sigma'}(\omega_n-\omega_n')\Bigr] \nonumber \\
&+\sum_{i_2} \mathcal{S}_{i_2,i_2+k}^{\sigma' \Lambda}(\omega_n') D_{ji_2\sigma\sigma'}^{(i-j)k\Lambda}(0)\Bigl\},
\end{align}
where the $l,k$-summation is restricted to $|l|,|k|\leq L$, whereas the sum over $i_2$ runs over the whole interacting region.
To summarize, dfRG2 is defined by the flow equations
\eqref{eq:matrixODE} and (\ref{eqSelfExplicit}), together with the
definitions \eqref{eqfRGgreen}, \eqref{eqStructures},
\eqref{eqNonzeroVertex}, \eqref{eq:crossfeed} and \eqref{eq:bubble}.

\subsection{Restrictions for actual computations}

In our actual computations, we restrict ourself to the case of zero temperature and use so called static fRG, meaning that we treat the vertices as frequency independent. The zero-temperature limit enables us to transform the summation over discrete Matsubara frequencies into continuous integrals along the imaginary axis, and the $\Theta_T$ in Eq.~\eqref{eqfRGscheme} is a sharp step function. Using this, we are able to apply Morris' lemma \cite{Morris1994}, which enables us to simplify the integral expressions containing the single-scale propagator $\mathcal{S}$ in the flow equations \eqref{eq:derive-flow-equations-channels-explicit}: under integration over $\omega$, the following relations hold: 

\begin{subequations}
\begin{align}
 \mathcal{S}^{\Lambda} (i \omega) &\stackrel{T=0}{=} \delta (|\omega|-\Lambda)
 \widetilde{\mathcal{G}}^{\Lambda} (i \omega), \\
 \widetilde{\mathcal{G}}^{\Lambda} (i \omega) &\; = \; \Big[ \left[
 \mathcal{G}_0 (i\omega )\right]^{-1} - \Sigma^{\Lambda} (i\omega) \Big]^{-1} ,
\\
 \mathcal{S}^{\Lambda}_{i,j} (i \omega_1^{~}) \mathcal{G}^{\Lambda}_{k,l} (i
 \omega_2^{~})
 &\stackrel{T=0}{=} \delta (|\omega_1^{~} |-\Lambda) \Theta ( |\omega_2^{~}
 |-\Lambda) \nonumber\\
  & \qquad \qquad \; \; 
 \widetilde{\mathcal{G}}^{\Lambda}_{i,j} (i \omega_1^{~})
 \widetilde{\mathcal{G}}^{\Lambda}_{k,l} (i \omega_2^{~})\; .
\end{align}
\end{subequations}

The static fRG approximation treats the vertex quantities
$\gamma_p^\Lambda$, $\gamma_x^\Lambda$, $\gamma_d^\Lambda$ as
frequency independent, setting the bosonic frequencies $\Pi$, $X$,
$\Delta$ to zero. Via Eq.~(\ref{eqgamma1DGL}), this automatically
implies that the self-energy is frequency independent, too. In the
case of QPC models with onsite interaction this approximation was
compared with results of the frequency dependent fRG-scheme, so called
``dynamic fRG'' and was seen to yield reasonable results for the
zero-frequency Green's function at zero temperature. However, for
models with finite-ranged interactions we find more pronounced static
fRG artifacts (described in section \ref{secFiniteInt}) which might be
improved by the use of the dynamical method. This is a topic for
future research. We stress here that it should in principle be
straightforward to implement the dynamical method. The main
restriction is simply the effort in computation time, which scales
like the number of used frequencies, $N_f$, which in 
Ref.~\onlinecite{Bauer2013} is typically of the order $10^2$.

\subsection{Numerical implementation}
In a numerical implementation, the flow will start at a value $\Lambda_i$ which is usually chosen as large, but is not infinite. For $\Lambda_i$ large enough, one can show \cite{Karrasch2006} that the flow of the self-energy from $\Lambda=\infty$ to $\Lambda=\Lambda_i$ results in a value of $\gamma_1^{\Lambda_i}$ given by

\begin{equation}
 \gamma_1^{\Lambda_i}
  ( \color{DGLorange} q_1' \color{black}, \color{RoyalBlue} q^{~}_1
   \color{black}) =
 -\frac{1}{2} \sum_{q} v (q ,\color{DGLorange} q_1' \color{black}; q,
  \color{RoyalBlue} q^{~}_1  \color{black}) \;  .
\end{equation}
This is then used as the initial condition for $\gamma_1$ in the numeric fRG flow. The initial condition for the vertex $\gamma_2$, given by Eq. (\ref{eq:initial-condition-vertex}), stays the same.

In the case of sfRG2 the vertices and the self-energy only depend on $\Lambda$. 
In order to carry out the resulting integration, we mapped the domain of the flow parameter $\Lambda \in [0,\infty)$ onto the finite domain $x \in [0,1)$ by using the substitution $\Lambda=\frac{x}{1-x}$, c.f. Ref.~\onlinecite{Bauer2014}. To integrate the resulting flow, we followed Dormand-Prince \cite{Dormand1980}, using a 4-th order Runge-Kutta method with adaptive step-size control.

\changed{For static fRG}, the computationally most expensive step is the block-matrix
multiplication of Eq.~\eqref{eq:matrixODE}, which scales as
$\mathcal{O} (N^3L^3)$. In \changed{dynamic fRG schemes with non-frequency cutoff (e.g. with
hybridisation flow \cite{Jakobs2010}), for intermediate
$N\lesssim10^2$ most of the calculation time is spent on the bubble integrals of Eq.~\eqref{eq:bubble}, whose
calculation time scales as $\mathcal{O} (N^2 L^2 N_f)$, where $N_f$ is the number of bosonic frequencies.
Since the numerical cost for this calculation (for the system sizes used in our setup) is comparable to the block-matrix multiplication of
Eq.~\eqref{eq:matrixODE}, it might be possible to implement the eCLA
within those schemes, too.}

\section{Results: Onsite-interactions} \label{secfRGresults}

Having derived our eCLA scheme in the last section, we are now able to apply it to the two models of primary interest here, namely the QPC and the QD. In the present section, we study purely onsite models  
\begin{equation}
 U_{ij} = \delta_{ij} U\, ,
\end{equation}
where we treat the strength $U$ of the interaction as a tunable and
space-independent parameter, which is suppressed smoothly to zero at
the ends of the interacting region.  The focus of this section lies on
comparing our results to the ones obtained previously by BHD to
explore the consequences of the improved feedback for a well-studied
example. \changed{If not otherwise specified}, plots in this section are calculated with $\mu=0$, i.e.\
with half-filled leads.

\begin{figure}
\includegraphics[width =83mm]{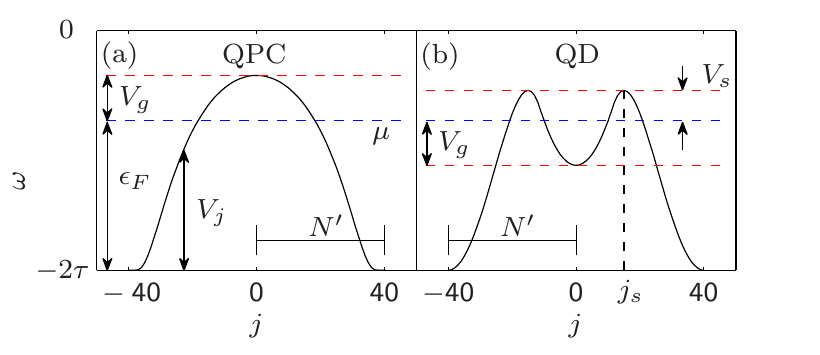}
\caption{Typical QPC and QD barrier shapes, controlled
  via the parameters, $\varepsilon_F$, $V_g$, $N'$,  and, 
  for the QD, $V_s$ and $j_s$. \changed{For these plots, 
both $\mu$ and the barrier top lie were chosen to lie below the center of the bulk band,
which we take as reference energy where $\omega = 0$. The case
of half-filled leads, used for most of our calculations,
corresponds to choosing $\mu = 0$.}
}
\label{Paper_barrieren}
\end{figure}

\subsection{Models for QPC and QD}
Our interest lies in the low energy physics of a QPC or a QD. For this
reason, we consider only the lowest subband of a QPC, or a QD coupled
to one-dimensional leads. We use a
  one-dimensional model Hamiltonian of the same form as used
in Refs.~\cite{Bauer2013,Bauer2014,Heyder2015}:
\begin{align}
\hat{H}\!=\! \sum_{j\sigma} [E_{j}^{\sigma} \hat{n}_{j\sigma}  -  \tau
 (d^\dagger_{j\sigma} d^{}_{j+1\sigma} + {\rm h.c.})] + \sum_j U_j
 \hat{n}_{j\uparrow}\hat{n}_{j\downarrow} . 
 \label{eqModelChain}
\end{align}
It describes an infinite tight-binding chain with constant lattice
spacing $a$, constant hopping amplitude $\tau$, on-site interaction
$U_j$, and on-site potential energy
$E_j^\sigma=V_j-\frac{\sigma B}{2}$. Here $V_j$ will be used to model
the smooth electrostatic QPC or QD potential defined by gates (as
described below and illustrated in Fig.~\ref{Paper_barrieren}), and
the Zeeman energy $B$ accounts for a uniform external magnetic field
parallel to the 2DEG. We take $U_j$ and $V_j$ to be nonzero only
within a (single or double) ``barrier region'' of $N = 2N'+ 1$ sites
centered around $j = 0$, containing the QPC or QD.  The rest of the
chain represents two noninteracting leads with bandwidth $4 \tau$,
chemical potential $\mu$, bulk Fermi energy
$\varepsilon_F = 2\tau + \mu$ and effective mass
$m^* = \hbar^2/(2 \tau a^2)$ (defined as the curvature of the
dispersion at the band bottom in the bulk).  Adopting the convention
in Ref.~\onlinecite{Bauer2014}, we choose the center of the bulk band
as energy origin.  
In order to arrive at a discrete QPC potential $V_j$, we start with a
continuous QPC potential
\begin{align}
V(x)=
\begin{cases}
(V_g + \varepsilon_F) \exp\left(\frac{-\gamma^2 (x/L_{\rm bar})^2}{1-(x/L_{\rm bar})^2}\right), & |x|\leq L_{\rm bar} \\
0 &  |x|> L_{\rm bar}
\end{cases}
\label{eqQPCcont}
\end{align}
where $2 L_{\rm bar}$ is the whole barrier length and $V_g$ controls the barrier height, measured
w.r.t.\ $\varepsilon_F$. Near the barrier top, the potential
\eqref{eqQPCcont} can be expanded as 
\begin{equation}
V(x)=V_g + \varepsilon_F - \frac{1}{2} \frac{m^*}{\hbar^2} \Omega_x^2 x^2 + \mathcal{O}(x^4) \, , 
\end{equation}
where the curvature parameter $\Omega_x$ is given by
\begin{align}
\label{eq:Omegax}
\Omega_x=\gamma\frac{\hbar }{L_{\rm bar}} \sqrt{\frac{2(V_g+\varepsilon_F)}{m^*}}.
\end{align} 
It has units of energy and serves as characteristic energy scale
for the QPC. It also defines a characteristic length scale for the
QPC barrier top
\begin{align}
\label{eq:lx}
l_x=\hbar/\sqrt{2m^*\Omega_x}=a\sqrt{\tau/\Omega_x}.
\end{align}
The dimensionless
parameter $\gamma$ in the exponent of \Eq{eqQPCcont} can be used to
vary the barrier curvature [\Eq{eq:Omegax}] without changing the
barrier height. Through most of section~\ref{secfRGresults}, we will
keep $\gamma=1$ constant and consider only gate-voltages small
compared to $\varepsilon_F$, such that the curvature can be assumed to
be independent of $V_g$. \changed{However, when studying eCLA convergence properties (Fig.~\ref{figCondLU_convergence}), and when 
dealing with longer-ranged
interactions in Sec.~\ref{secFiniteInt}, we will need to choose
$\gamma \neq 1$}.

We discretize the QPC potential \eqref{eqQPCcont} by choosing
a number of sites $N$ and setting the lattice spacing $a=2 L_{\rm
  bar}/N$, to arrive at 
\begin{equation}
V_j=V(j\cdot a)=
\begin{cases}
(V_g+\varepsilon_F) e^{-\gamma^2\frac{(j/N')^2}{1-(j/N')^2}}, &\ |j|\leq N' \, , \\ 
0, &\ |j|>N' \, . 
\end{cases}
\label{eqPotentialDiscrete}
\end{equation}
The resulting barrier shape given by Eq.~\eqref{eqPotentialDiscrete} is plotted in Fig.~\ref{Paper_barrieren}(a). The leading behavior around the maximum at $j=0$ is quadratic and the same as in Ref.~\onlinecite{Bauer2014}:
\begin{equation}
V_j= V_g+\varepsilon_F-\frac{\Omega_x^2}{4\tau} j^2+\mathcal{O}(j^4),
\end{equation}
and the curvature can be expressed through the discrete quantities as
$\Omega_x=\gamma \frac{2\sqrt{\tau (\varepsilon_F+V_g)}}{N'}$.
For our onsite studies, where $V_g$ is only varied in a small
region around $V_g=0$, we use the approximation
$\Omega_x=\gamma \frac{2\sqrt{\tau \varepsilon_F}}{N'}$. In order to
avoid discretization artifacts, the discretization length $a$ should
be chosen significantly smaller than $l_x$. In our actual computations
for the QPC with onsite interactions we use a \changed{ratio $l_x/a$ varying between approximately $4-10$ sites.}

To model a QD, we use a potential that can be tuned smoothly from the
QPC shape described above to a double-barrier structure, as shown in
Fig.~\ref{Paper_barrieren}(b). The discretization procedure is
analogous to the QPC and we state here only the resulting discrete dot
potential, which is the same as used in
Refs.~\onlinecite{Bauer2013,Heyder2015}:

\begin{eqnarray} \nonumber &V_j \,&\!=\begin{cases}
    0 \; , & \hspace{-1.cm}  \forall \; |j| \ge N'  ,  \vspace{3mm} \\
    (V_s+ \varepsilon_F) \left[2 \left(\frac{|j|- N'}{j_s- N'}\right)^2
        - \left(\frac{|j|- N'}{j_s- N'}\right)^4 \right] , \quad
    & \\
    & \hspace{-1.7cm} \forall  \;  j_0 \leq |j| \le  N' , \vspace{3mm}  \\
    V_g + \varepsilon_F + \frac{\bar{\Omega}_x^2 j^2}{4 \tau} \, {\rm sgn}(V_s
    - V_g) , & \hspace{-1.5cm} \forall \; 0 \leq |j| < j_0 .
\end{cases}\\
\label{eq:potential}
\end{eqnarray}
We can vary the dot width via $j_s$, and the depth of the quadratic well in the middle via $V_s$ and $V_g$. These choices determine the values of $j_0$ and $\bar{\Omega}_x$ in order to make the potential continuously differentiable. Of course, this is just one convenient way to model the dot structure, and the qualitative behavior of the physical results does not depend on the specific implementation.

For the onsite interaction we use both for the QPC and the QD the form
used by BHD \cite{Bauer2013}:
\begin{align}
U_j=U e^{- (j/N')^6/[1-(j/N')^2]}.
\label{eq:interaction}
\end{align}
It is almost constant and equal to $U$ in the center of the QPC and
drops smoothly to zero at the flanks of the barrier region.

\subsection{Physical behavior of the models}
We now briefly summarize the physics of these models, which was
already discussed in great detail by BHD in
Refs.~\onlinecite{Bauer2013,Heyder2015}.  Our main handle for tuning
the QPC potential is the gate voltage $V_g$, which controls the height
of the barrier. If the barrier top lies well above the chemical
potential, the QPC is closed. Lowering the barrier, the QPC opens up
and the linear conductance $g$ increases smoothly from 0
 to 1 in the region of gate voltages $0 \lesssim
V_g \lesssim \Omega_x$, where $\Omega_x$ 
is the curvature of the QPC introduced above. Additionally, the
width of the conductance step, i.e.\ the gate-voltage interval in
which the conductance increases from zero to one, is also set by
$\Omega_x$. The general shape of the conductance curve for a
parabolic barrier in the absence of interactions is a step described
by a Fermi-function, as was shown by B\"uttiker in
Ref.~\onlinecite{Buttiker1990}. If one switches on onsite
interactions, the conductance curve becomes asymmetric and flattens
increasingly at the top. This effect can be traced back to the fact
that when the barrier top drops below the chemical potential as the
QPC is being opened up, the maximum in the LDOS just above the
barrier top (called van Hove ridge in Ref.~\cite{Bauer2013}) is
aligned with the chemical potential, thereby strongly enhancing
interaction effects. It turns out that the effective onsite
interaction strength is in fact given by
\begin{equation}
 U^\text{eff }_j=U \cdot \mathcal{A}^0_{j}(\mu),
\label{Ueff}
\end{equation}
where
\begin{align}
\mathcal{A}^0_j(\omega) = -\frac{1}{\pi} {\rm Im} \mathcal{G}_{jj}^0 (\omega + i 0^+)
\end{align}
is the non-interacting local density of states per site. Near the barrier center, the resulting $U^\text{eff}$ scales like $U/\sqrt{\Omega_x \tau}$.
 
In the QD case, we can vary the width and depth of the middle well,
[c.f.\ Fig.~\ref{Paper_dot}~(d,e) below]. Typically, we want to study
the crossover between QPC and QD, thus we start out with a QPC setup
and lower the potential of the central region to change the geometry
to a QD model. The characteristic physics of the quantum dot is
determined by the structure of the discrete levels of the bound states
in the well. This quantization leads to a conductance peak whenever
such a level crosses the chemical potential and the dot gets filled by
one electron more. In the interacting case, the degenerate levels
split on a scale of the interaction strength $U$. However, there is a
further effect: the odd valleys, i.e.\ the regions between the peaks
where the dot contains an odd number of electrons, become conductance
plateaus with $G_Q\approx 1$. This behavior reflects the occurrence of
the Kondo \cite{Kondo1964} effect since the singly-occupied dot level
behaves like a localized spin coupled to a fermionic bath.

In this work, we will apply our eCLA first to the same type of onsite
models of QPCs as used by BHD \cite{Bauer2013,Bauer2014,Heyder2015} and
analyse the resulting effects. Importantly, we find that in comparison
to the CLA used previously, the eCLA yields an improved stability of
the fRG flow in the case of large bare LDOS at the chemical
potential. This improvement allows us to additionally study the QPC-QD
crossover, which involves a very high LDOS due to the flat barrier top
that occurs in this transition. Using the CLA, it had not been
possible to study this transition when the barrier top lies close to
the chemical potential $\mu$, since the CLA equations did not
converge. Due to this problem, in the real-space approach chosen by
Heyder \textit{et al.}  \cite{Heyder2015} it was not possible to study
dots which contain just a few electrons. Since our new feedback scheme
significantly ameliorates the convergence problem, we are now able to
study the crossover from a QPC to a QD which is just occupied by a
single electron. This will be shown in section
\ref{Transition_QPC_QD}.

\begin{figure*}
  \includegraphics[width
  =183mm]{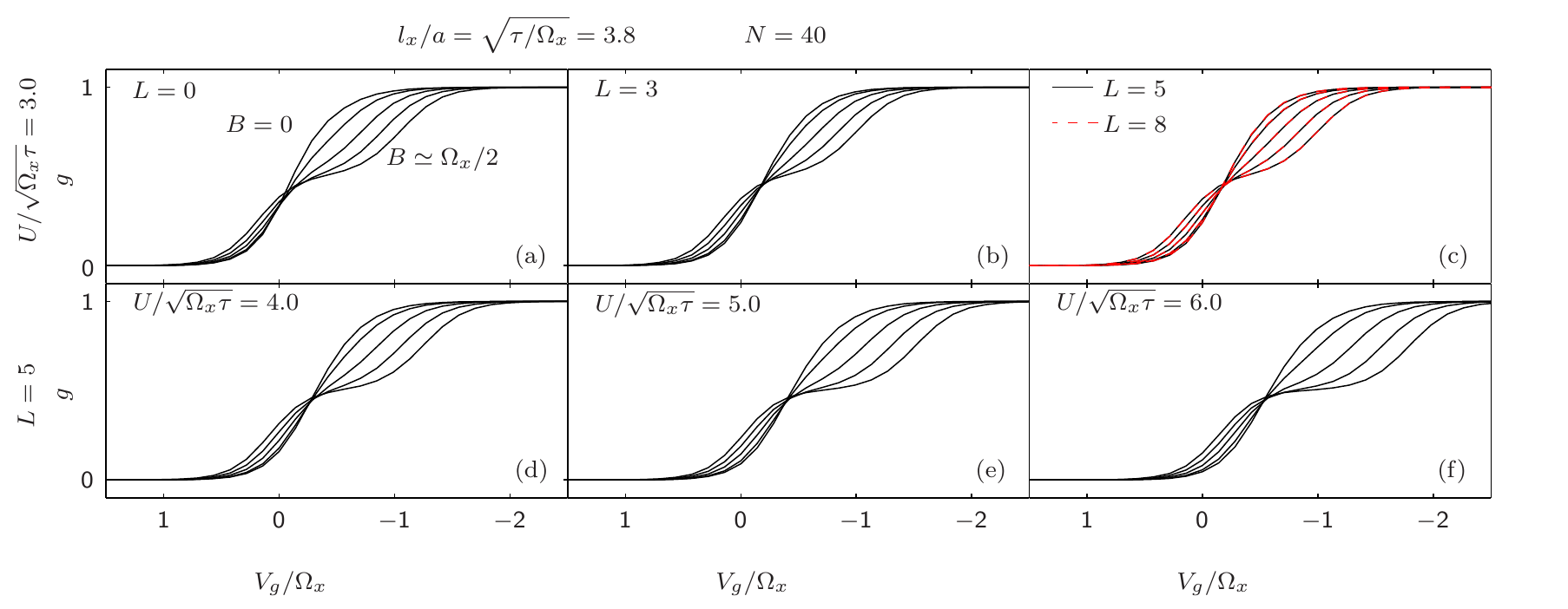} \caption{Linear conductance $g$
    calculated using the static eCLA for five equidistantly chosen
    magnetic fields $B$ between $0$ and $\Omega_x/2$.  (a-c)
    Conductance at fixed $U/\sqrt{\Omega_x \tau}=3.0$, and four values
    of $L$.  (d-f) Conductance at fixed $L=5$, for three values of
    $U/\sqrt{\Omega_x \tau}$.}
\label{figCondLU}
\end{figure*}

\subsection{Increasing the feedback length}
\label{secCond}

Let us now study the influence of the feedback length $L$ on the zero-temperature linear conductance \cite{Datta1997},

\begin{equation}
\label{eqcond}
 g=\frac{1}{2} \sum_\sigma \left| 2\pi \rho^{\sigma}(\mu + i0^+)
 \mathcal{G}_{-N'N'}^{\sigma} (\mu + i0^+) \right|^2 \, .
\end{equation}
Here $\rho(\omega)$ is the density of states at the boundary of a
semi-infinite tight-binding chain; two such chains represent the two
one-dimensional non-interacting leads, coupled to the central
interacting region.  Let us first look at the QPC case. We are
interested in the shape of the conductance trace as a function of
applied gate voltage and how this shape changes with external
parameters, such as applied magnetic field.

For pure onsite interactions, it is natural to choose the feedback
length $L=0$. This is what has been done in
Refs.~\onlinecite{Bauer2013,Bauer2014,Goulko2014,Heyder2015}, and the
results have been discussed therein in detail. Here, we will allow a
nonzero $L$, although the actual interaction is purely onsite.  This
implies that a certain class of additional third order terms will be
generated during the RG flow which introduce a better coupling between
the channels in the sense of the feedback in
Eq.~\eqref{eqFeedback}. For $L\to N$ the third-order truncated static
fRG scheme is recovered \textit{fully} regarding the spatial
structure of the two-particle vertex (but not for its frequency
structure, since we are using the static approximation).
Figs.~\ref{figCondLU}(a)~to~(c) show the conductance $G$ as a
function of gate voltage $V_{\rm g}$ for different values of
magnetic field $B$, calculated at fixed $U$ and different values of
feedback parameters $L$. Increasing the latter from $L=0$ to $L=3$,
c.f.\ Fig.~\ref{figCondLU}(b), leads to quantitative but not
qualitative changes in the shape of the conductance curves -- the
main effect is that the width of the $B$-induced subplateau 
decreases. In this regard, increasing $L$ has a qualitatively
similar effect to decreasing $U$ (at $L=0$), c.f.\
Fig.~\ref{figCondLU}(d)~to~(f).  Note, though, that increasing $L$
hardly affects the $V_g$ position of the conductance step, whereas
decreasing $U$ does shift the step slightly towards higher $V_g$
values, as expected physically due to the lowering of the Hartree
barrier.  Increasing the feedback beyond $L=5$ does not lead to any
significant quantitative changes, as can be seen in
Fig.~\ref{figCondLU}(c) where $L=5$ (black line) is directly
compared with $L=8$ (red dashed line).  Hence, for the present model
convergence is reached for\changed{ $L\lesssim5$.} In general this value depends on
the strength of interaction $U$, and more importantly on the actual
shape of the barrier.

\begin{figure}
  \includegraphics[width
  =83mm]{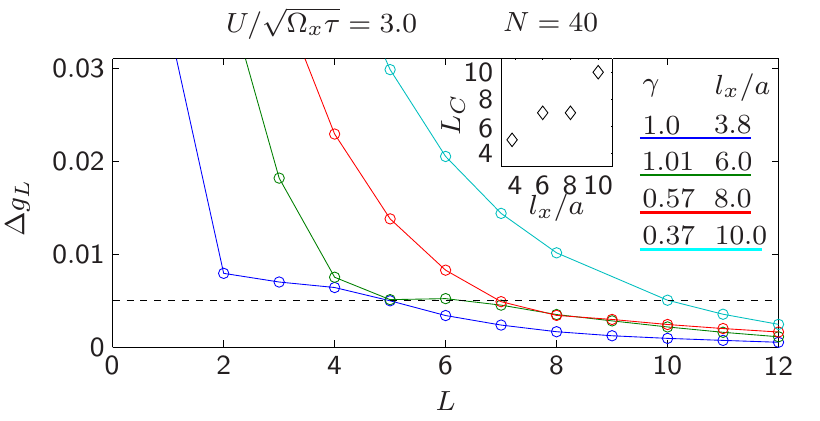} \caption{Convergence
    behavior of the conductance for different values of $l_x/a$, where
    $\Delta g_L$ is defined in Eq.~\eqref{eq:deltagL}. The parameters
    for the $l_x/a=3.8$ data are the same as in
    Fig.~\ref{figCondLU}. For the larger $l_x$ values the chemical
    potential was chosen as $\mu=-1.7$ and the parameter $\gamma$ was
    varied. The inset shows the dependence of $L_C$ on $l_x$. }
\label{figCondLU_convergence}
\end{figure}

\changed{In Fig.~\ref{figCondLU_convergence} we study the convergence behavior as function of the feedback length $L$ more thoroughly, 
for four different values of the geometric length scale $l_x/a$ [Eq.~\eqref{eq:lx}] which is the width of the region where the LDOS is enhanced.
To determine the convergence behavior, we first chose a large value $L_{\rm large}$ (here $L_{\rm large}=21$) for which $\max_{V_g} |g_{L_{\rm large}}(V_g) -g_{L_{\rm large}-1}(V_g) | $ is smaller than $10^{-4}$, i.e for which we can assume that the conductance is converged against its limit. We then plot
\begin{align}
\Delta g_L:= \max_{V_g} |g_{L}(V_g) -g_{L_{\rm large}}(V_g) | 
\label{eq:deltagL}
\end{align}
as a function of $L$. For our purposes, as for the plots in Fig.~\ref{figCondLU}, we will regard that the conductance as being converged when $\Delta g_L\leq 0.5 \cdot 10^{-2} $. In  Fig.~\ref{figCondLU_convergence} this criterium is indicated by the dashed line. The inset shows the smallest $L$ (named $L_C$) for which the conductance is converged as a function of $l_x/a$.
We see that for all models under our consideration $L_C$ is comparable to
$l_x/a$. 
Due to this convergence, the number of vertex components can safely be reduced from $\mathcal{O}(N^4)$ to $\mathcal{O}(N^2 L^2)$, where $L \approx l_x/a$. It would be interesting to investigate if this number can be reduced even further, a next possible candidate being $\mathcal{O}(N L^3)$, by studying the structure of the vertex in more detail. This is, however, beyond the scope of this work and we leave this question for further research.}

The extended feedback between the channels becomes increasingly
important with increasing interaction strength.  For $L=5$ the eCLA
yields meaningful, converged results for interaction values for
which the $L=0$ flow obtained by CLA is divergent. This is the case
for $U\gtrsim4\sqrt{\Omega_x\tau}$.  Figs.~\ref{figCondLU}(d)~to~(f)
show the conductance for such large values of interaction and
$L=5$. The qualitative behaviour is unchanged w.r.t.\ smaller values
of the interaction, and the quantitative strength of the impact of
the interaction increases continuously, in that the width of the
spin-split subplateau increases with $U$.

\begin{figure}
\includegraphics[width =83mm]{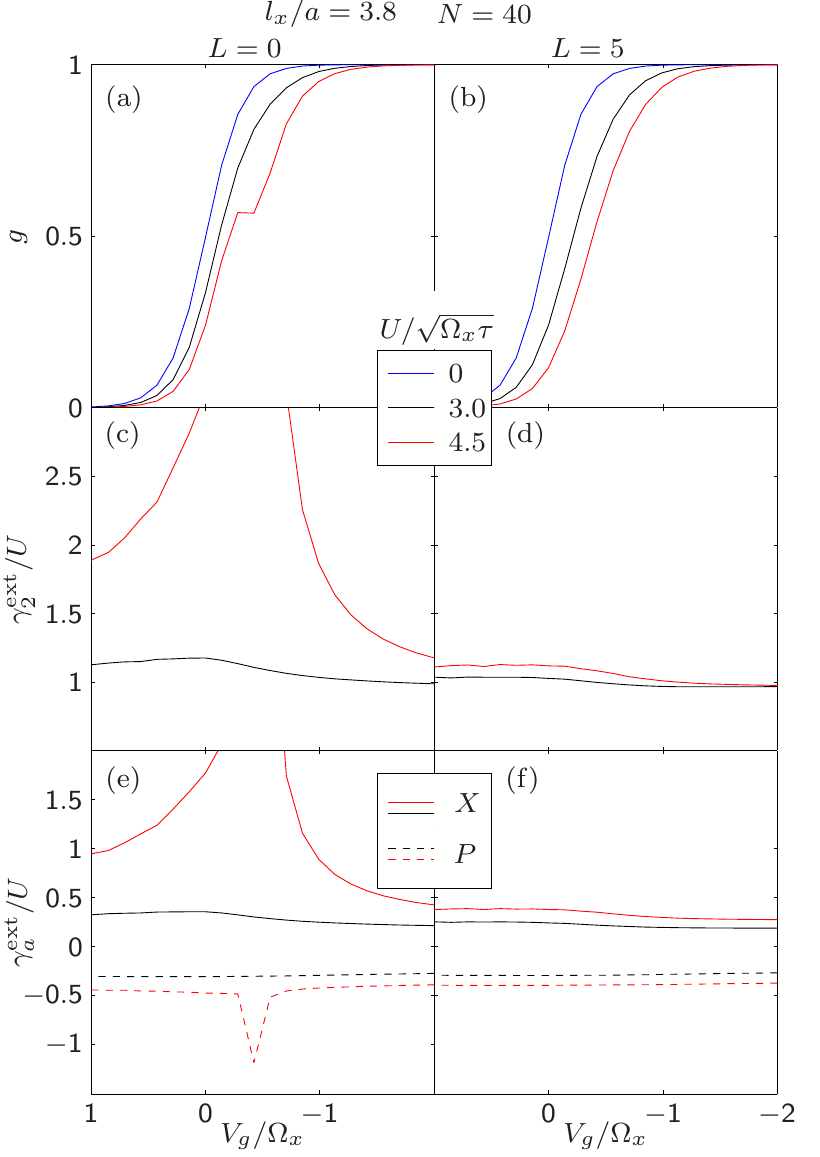}
 \caption{Conductance and vertex quantities calculated for the two feedback lengths $L=0$ (left column) and $L=5$ (right column) with three different effective interaction strengths $U/\sqrt{\Omega_x\tau}$, at zero magnetic field.} 
\label{Paper_analysis_vertex3}
\end{figure}

To shed light on the effect of the enhanced coupling
between the channels, we now analyse the resulting two-particle vertex quantitatively,
by studying its extremal value
\begin{align}
 \gamma_2^{\rm ext} &= 
 \max_{q_1' q_2' q_1^{} q_2^{}} |\gamma_2(q_1',q_2';q_1^{},q_2^{})|,
 \nonumber \\
\end{align}
where the $q$'s stand here both for site and spin indices. Furthermore, we identify the two most contributing parts to these value as
\begin{align}
 \gamma_x^{\rm ext} &= 
 \max_{j_1' j_2' j_1^{} j_2^{}} \gamma_x(j_1' \uparrow,j_2' \downarrow;j_1^{}\uparrow,j_2^{}\downarrow)
 \nonumber \\
 \gamma_p^{\rm ext} &= 
 \min_{j_1' j_2' j_1^{} j_2^{}} \gamma_p(j_1'\uparrow,j_2'\downarrow;j_1^{}\uparrow,j_2^{}\downarrow)
 \, .
\end{align}
Note that we used the minimum in the definition of $\gamma_p^{\rm ext}$, since the $\gamma_p$ contribution is mainly negative, whereas $\gamma_x$ is dominated by its positive part. Fig.~\ref{Paper_analysis_vertex3} shows these quantities
and the conductance  as a function of $V_g$ for $L=0$
and $L=5$.
The main message of this figure is that for intermediate interaction
strength (solid black curves) the flow converges for both $L=0$ (left
column) and $L=5$ (right column) and yields qualitatively the same
results for the conductance in
Fig.~\ref{Paper_analysis_vertex3}~(a,b). If, however, one increases
the interaction strength further (red solid curves) the flow for $L=0$
starts to diverge [Fig.~\ref{Paper_analysis_vertex3}~(c,e)] and the
values of physical observables computed from it become wrong,
reflected for example in the kink of the red conductance curve in
Fig.~\ref{Paper_analysis_vertex3}~(a). A good measure for the behavior
of the flow is the maximum value of the two-particle vertex, plotted
in Fig.~\ref{Paper_analysis_vertex3}~(c,d). We see that the kink in
the conductance curve corresponds to a very large value of
$\gamma_2^{\rm ext}/U=58.2$ [lying outside of the range of
Fig.~\ref{Paper_analysis_vertex3}(c)]. In contrast, for $L=5$,
$\gamma_2^{\rm ext}$ as well as the conductance stay well behaved and,
in fact, the flow converges without problems
[Fig.~\ref{Paper_analysis_vertex3}~(b,d)]. In order to shed light on
this stabilizing effect of the enhanced feedback, we show in
Fig.~\ref{Paper_analysis_vertex3}~(e,f) the $P^{\uparrow \downarrow}$
and $X^{\uparrow \downarrow}$ part of the channels, which constitute
the contributions to $\gamma_2^{\rm ext}$ with the largest moduli. In
the case of intermediate interaction (black curves) the $X$ and $P$
contributions are of the same order of magnitude but differ in their
relative sign. If one looks at the completely uncoupled channels,
i.e.\ the pure ladder contributions (c.f.\ the study in
Ref.~\onlinecite{Bauer2014}) and increases the interaction strength,
the X-channel is the first one to diverge. Our interpretation of the
stabilizing effect is now as follows. Since the channels are coupled,
a slight increase in the modulus of the X-channel leads via the
feedback to a slight increase of the modulus of the P-channel, and due
to their relative sign difference they partially cancel, so that the
resulting additional contribution to $\gamma_2$ is small. If the
effective interaction becomes too strong, this ameliorating effect
eventually breaks down and the flow diverges. In the $L=5$ case, we
take much more feedback between the individual channels into account
than for $L=0$ and it is therefore reasonable that the divergence
point of the flow is shifted toward larger effective interactions.
 
\subsection{Crossover between a closed QPC and a QD}
\label{Transition_QPC_QD}
As we have seen above, the increase of the feedback length $L$ leads
to a more stable fRG flow in regions for high LDOS, corresponding to a
large effective interaction strength. This stabilization effect
enables us to study parameter regimes that have been \changed{hard to
  treat with} previous fRG schemes. We illustrate this below for a
situation known to suffer from fRG divergence problems, namely the
crossover from a QPC to a QD. \changed{In Ref.~\onlinecite{Heyder2015}
  it was found that when using the CLA (called ``fRG2'' there), the
  fRG flow for this transition suffers from divergences if the flat
  barrier top is too close to the chemical potential. For this reason,
  it was not possible for fRG2 to smoothly describe how the dot
  filling increases with decreasing $V_g$, and the region where no or
  only a few electrons occupy the dot remained inaccessible within the
  CLA.
  The eCLA enables us now to study precisely this interesting region.
  (In Ref.~\onlinecite{Heyder2015}, this regime was treated instead
  using a simpler fRG scheme without vertex flow (``fRG1''). Although
  this did qualitatively produce the Kondo physics that is expected if
  the QD occupancy is odd, Ref.~\onlinecite{Heyder2015} argued that
  fRG1 is generically less reliable than fRG2. For example, for a QPC
  geometry, it underestimates the scewing of the zero-temperature
  conductance step that is characteristic for the 0.7-anomaly. For
  this reason, the detailed studies of QD-QPC crossovers performed in
  Ref.~\onlinecite{Heyder2015} were all limited to deep dots, studied
  using fRG2.)}

\begin{figure*}
 \includegraphics[width =183mm]{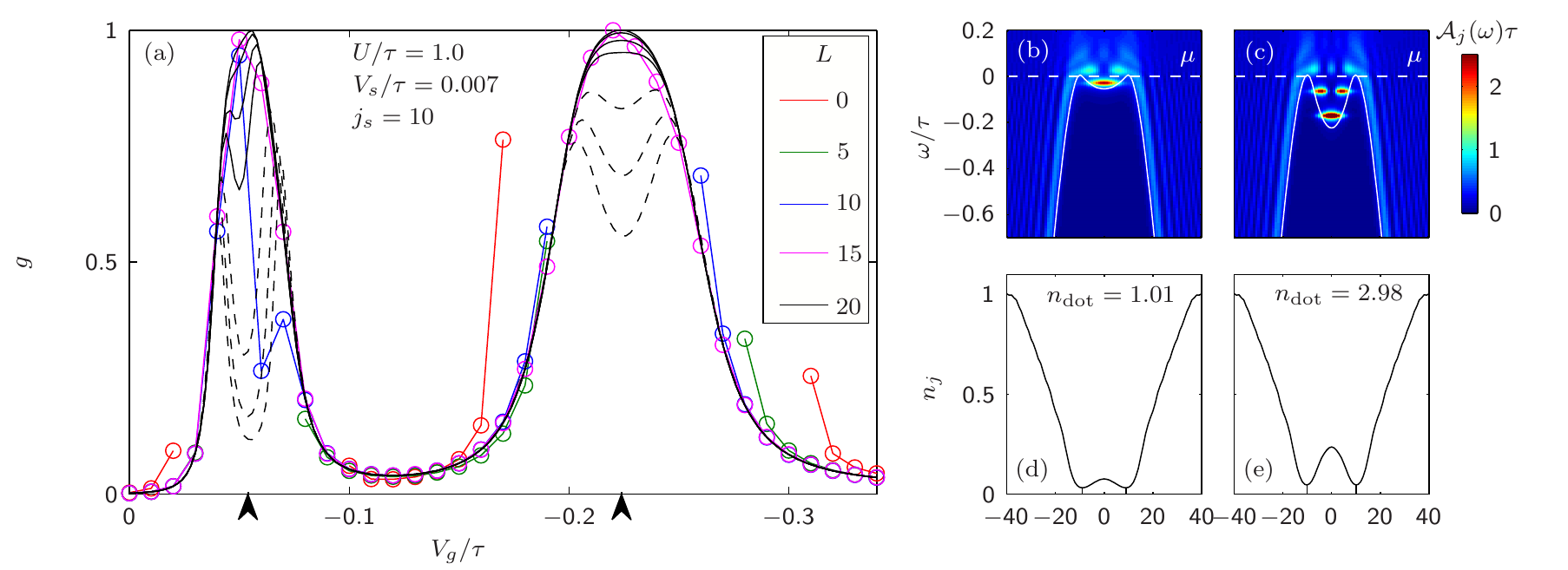}
 \caption{\label{Paper_dot}The crossover from a QPC to
   a QD. (a) The conductance as function of gate-voltage $V_g$,
   calculated for several magnetic fields (black solid lines:
   $B=0,1,2,3\cdot 10^{-4}$, black dashed lines:
   $B=6,9,12\cdot 10^{-4}$) with feedback-length $L=20$. Colored
   symbols indicate the conductance values obtained with smaller
   feedback lengths.
   \changed{  (b), (c) Non-interacting LDOS (color scale) and
   barrier shape (solid white curve) for the two gate voltages marked by the left and right
   vertical arrows in (a), respectively.
   Horizontal white dashed lines indicate
   the chemical potential $\mu$. 
  (d), (e) The electron density per site, $n_j$,
   again computed for the two gate voltages indicated in 
   (a). Summing $n_j$ over all sites between the two density
     minima yields $n_{\rm dot}=1.01$ and $n_{\rm dot}=2.98$.}
   }
\end{figure*}

\changed{In Fig.~\ref{Paper_dot}~(a)  we show the conductance curve for the crossover between a closed QPC and a QD, in which the first two bound state levels cross the chemical potential as the dot is made deeper. This level structure is illustrated in Fig.~\ref{Paper_dot}~(b,c) where we show the noninteracting LDOS of the dot structure for the two gate voltages indicated by the black markers in  Fig.~\ref{Paper_dot}~(a). Both of these gate voltages lie within regions where the sharp LDOS maximum associated with a bound state near $\omega=\mu$ causes convergence problems if the feedback length $L$ is small, but not if $L$ is chosen sufficiently large, which is possible within the eCLA.}

\changed{When varying gate voltage, we can see Kondo plateaus in the conductance arising in the $V_g$ regions where the occupation of the dot is odd. This is illustrated in Fig.~\ref{Paper_dot}~(d,e), where we show the site-resolved density, again for the two $V_g$ values indicated in (a). We see that the electrons are localized within the QPC, which here had a width of $20$ sites. When the densities within the QPC are integrated, we indeed obtain approximately one electron for the first plateau and three electrons for the second plateau.

These Kondo plateaus, caused by Kondo screening of the dot spin, get suppressed with increasing magnetic field since the spin degeneracy is broken.}
This suppression happens in the first and second Kondo plateau for magnetic fields on the scale $\sim 1\cdot10^{-4}\tau$ (solid black lines), and $\sim 3\cdot 10^{-4}\tau$ (dashed black lines), respectively.
A quantitative extraction and analysis of the Kondo scales of the setup is beyond the scope of this paper.
Our main purpose here is to illustrate that the finite-ranged feedback
of eCLA enables us to treat a parameter regime which was not accesible
with previous fRG schemes and produces qualitatively correct Kondo
physics. To outline this, we have indicated in
Fig.~\ref{Paper_dot}~(a) how the range of convergence increases with
increasing $L$ from $0$ to $30$. We see that the $L=0$ method is only
convergent in the parameter regimes where the occupancy of the dot is
even and hence the conductance is small. By increasing $L$ from $0$
over $5$ to $10$, we see that also the conductance plateaus become
more and more visible. At $L=20$ the whole Kondo plateau is accesible.
Upon further increasing the feedback up to $L=30$ (not shown here), we
find that the conductance results for $L=20$ are already properly
converged.

\section{Finite-ranged interactions} \label{secFiniteInt}
\label{Longer-ranged interactions}

In this section we consider a model of a QPC with an interaction
whose range extends over up to $N$ sites, in contrast to the purely
onsite interaction studied in Sec.~\ref{secfRGresults}. The purpose
of this study is to illustrate the potential of the eCLA to deal
with finite-ranged interactions in a setting where the screening of
a longer-ranged interaction comes into play, and to take a first
step towards exploring the physical consequences of screening. We
should emphasize, though, that we do \textit{not} aim here to
achieve a fully realistic treatment of screening in a QPC. That
would require including higher-lying transport modes (we consider
just the lowest-lying one), which would go well beyond the scope of
the present paper.

Our model is described by the following Hamiltonian:
\begin{align}
\begin{split}
  \hat{H}=&\sum_{ij\sigma} [E_{j}^{\sigma} \hat{n}_{j\sigma} \! - \! \tau
 (d^\dagger_{j\sigma} d^{}_{j+1\sigma} + {\rm h.c.})] \\
 &+ \frac{1}{2}\sum_{i,j,\sigma,\sigma'} U_{ij} 
 \hat{n}_{i\sigma}\hat{n}_{j\sigma'}(1-\delta_{ij}\delta_{\sigma \sigma'}). 
 \label{eqModelChain2}
\end{split}
\end{align}
Here $E_j^\sigma$ is chosen as described in section
\ref{secfRGresults}, and $U_{ij}$ can differ from zero for all sites
with separation $|i-j|<L_U$, where $L_U$ determines the bare
interaction range. Note that we now also have a bare interaction
between electrons with the same spin, which was absent in the onsite
case. In the previous section, the interaction strength was
controlled by a single value $U$ [c.f.\ eq.~\eqref{eq:interaction}]
and treated as a tunable parameter, whose strength was varied by
hand. However, now $U_{ij}$ is a matrix with $N^2$ parameters, and we
need to specify its form explicitly.  For this we start with a
continuous 3D model of a QPC, and for the Hilbert spaces associated
with transverse motion in the $y$- and $z$-directions we reduce the
dimensionality down to one, by taking into account only the ground
states of the respective confining potentials, c.f.\
Ref.~\onlinecite{Lunde2009}.
In this way, we arrive at a continuous effective theory in 1D for the
$x$-direction, which in a last step is discretized using a finite
difference method, already applied by BHD in
Ref.~\onlinecite{Bauer2014}.  We use the resulting model to compute
the conductance and the density profile of a QPC, and study their
dependence on the screening effects of the long-ranged interaction
and the geometric dimensions of the QPC.

\subsection{Derivation of a 1-dimensional Hamiltonian}
We start from the Hamiltonian $\hat{H}=\hat{H}_0+\hat{H}_1$ with
\begin{equation}
\begin{split}
 \hat{H}_0=&\sum_\sigma \int d^3r \hat{\Psi}^\dagger_\sigma(\mathbf{r})\left(V_{\rm QPC}(\mathbf{r})-\frac{\hbar^2}{2m}\nabla^2\right)\hat{\Psi}_\sigma(\mathbf{r})\\
 \hat{H}_1=&\frac{1}{2}\sum_{\sigma_1,\sigma_2}\int d^3 r_1 \int d^3 r_2 U(\mathbf{r}_1-\mathbf{r}_2) \\
     &\times\hat{\Psi}^\dagger_{\sigma_1}(\mathbf{r}_1)\hat{\Psi}^\dagger_{\sigma_2}(\mathbf{r}_2)\hat{\Psi}_{\sigma_2}(\mathbf{r}_2)\hat{\Psi}_{\sigma_1}(\mathbf{r}_1),
\label{eqModelChain2b}
\end{split}
\end{equation}
where the fermionic field $\hat{\Psi}^\dagger_\sigma(\mathbf{r})$
creates an electron with spin $\sigma$ at the continuous position variable
$\mathbf{r}$. The interaction is of screened Coulomb form with
screening length $l_s$ and relative dielectric constant $\kappa$,
which is given in ESU-CGS units by:
\begin{equation}
U(\mathbf{r}_1-\mathbf{r}_2)=\frac{e^2}{\kappa} \left( \frac{1}{|\mathbf{r}_1-\mathbf{r}_2|}-\frac{1}{\sqrt{|\mathbf{r}_1-\mathbf{r}_2|^2+l_s^2}} \right),
\end{equation}
c.f.\ Hirose \textit{et al.}  \cite{Hirose2003}.
This interaction form results from taking image charges on the top gate into account, which is positioned at a distance of $l_s/2$ above the 2DEG.
We use a QPC potential given by
\begin{equation}
 V_{\rm QPC}(x,y,z)=\left[\alpha V(x)+m^*\frac{\Omega_y(x)^2}{\hbar^2} \frac{y^2}{2}\right] \Delta(z),
\label{V_QPC}
\end{equation}
with $\Omega_y(x)=2 \beta V(x)$, and $m^*=0.067m_e$ is the effective
mass of GaAs.  The function $\Delta(z)$ ensures the confinement to the
2DEG and the one-dimensional potential $V(x)$ which enters here is the
same as that used in our onsite-model studies, Eq.~\eqref{eqQPCcont}.
The QPC potential $V_{\rm QPC}$ has a saddle-like form: it defines a
quadratic confinement in $y$-direction with a positiv curvature
$\Omega_y(x)$ that decreases with increasing $|x|$, whereas the
curvature in $x$-direction is negative, with magnitude $\Omega_x$. The
confinement in $y$-direction disappears for $|x|\to \infty$, where
$V(x)=0$.
For the coefficients $\alpha$ and $\beta$, we impose the condition
$\alpha+\beta=1$, which turns out to ensure that the effective
one-dimensional potential resulting from eliminating the $y$- and
$z$-direction is precisely $V(x)$.  We specify the transverse
curvature at the center of the QPC to be $\Omega_y=\Omega_y(0)$, thereby 
fixing the parameter $\beta=\frac{\Omega_y}{2 V(0)}$.

We now project onto the ground state subspace for the transverse
directions. With this step, taken for the 
sake of simplicity, we ignore all transport
modes except the one contributing to the first conductance step.
For a truly realistic description of screening, 
the higher-lying modes would have to be taken
into account, too. This would lead to stronger screening
and an effective interaction of shorter range than that
obtained below. 

Concretely, we thus represent our quantized fields as
\begin{align}
\hat{\Psi}_\sigma(\mathbf{r})= \phi_x(y) \varphi (z) \hat{\psi}_\sigma (x) .
\end{align}
Here $\phi_x(y)$ and $\varphi(z)$ are the normalized ground state
wave-functions of the confining potentials in the $y$ and $z$
directions, respectively ,
\begin{align}
 \varphi(z)&=\sqrt{\tilde{\delta}(z)}, \\
 \phi_x(y)&=\frac{1}{(2\pi)^{1/4}\sqrt{l_y(x)}}e^{-y^2/(4l_y^2(x))}
\end{align}
and the operator $\hat{\psi}_\sigma(x)$ creates an electron in a state
with wavefunction $\delta(x) \phi_x(y) \varphi(z)$.  In our 2DEG
setup, $\tilde{\delta}(z)$ is a peak of weight one, very narrow
compared to the scales in $x$- and $y$-direction, whereas
$\phi_{x}(y)$ is the ground state of a harmonic oscillator with
characteristic length
\begin{equation}
l_y(x)=\frac{\hbar}{\sqrt{2m^* \Omega_y(x)}}.
\end{equation}
With this, we arrive at an effective 1D continuous theory described by the effective 1D Hamiltonian 
\begin{equation}
\begin{split}
 &\hat{H}_{\text{eff}}=\sum_{\sigma}\int dx \hat{\psi}_\sigma^\dagger(x)\left[\frac{\hbar}{2m}\partial_x^2+ (\alpha+\beta) V(x)\right] \hat{\psi}_\sigma(x)\\
 &+\!\!\sum_{\sigma_1, \sigma_2}\!\! \int\!\! 
 dx_1 dx_2 \frac{U(x_1,x_2)}{2} 
\hat{\psi}_{\sigma_1}^\dagger\! (x_1)\hat{\psi}_{\sigma_2}^\dagger\! (x_2)
\hat{\psi}_{\sigma_2}\! (x_2)\hat{\psi}_{\sigma_1}\! (x_1).
\end{split}
\label{eqEffectiveHamiltonian}
\end{equation}
We now choose $\alpha+\beta=1$ as stated above, thus ensuring that
the resulting effective one-dimensional potential is indeed given by
$V(x)$.
The matrix elements of the interaction are given by
\begin{equation}
 \begin{split}
  &U(x_1,x_2)=
  \frac{e^2}{\kappa} \sqrt{\frac{1}{2\pi(l_y^2(x_1)+l_y^2(x_2))}}\\
  &\times \left[ \exp\left(\frac{(x_1-x_2)^2}{4(l_y^2(x_1)+l_y^2(x_2))}\right) \cdot K_0\left(\frac{(x_1-x_2)^2}{4(l_y^2(x_1)+l_y^2(x_2))}\right)\right. \\
  &-\left.\exp\left(\frac{(x_1-x_2)^2+l_s^2}{4(l_y^2(x_1)+l_y^2(x_2))}\right) \cdot K_0\left(\frac{(x_1-x_2)^2+l_s^2}{4(l_y^2(x_1)+l_y^2(x_2))}\right)\right].
\end{split}
\label{interaction_1d}
\end{equation}
For a typical 2DEG of GaAs-AlGaAs the relative dielectric constant has the value $\kappa \approx 12.9$. $K_0$ is the modified Bessel function of second kind in zeroth order. It diverges logarithmically when its argument approaches zero.

In order to discretize our 1D continuous theory along the
$x$-direction, we set $x:=a \cdot j$ and replace the continuous field
$\hat{\psi}_\sigma(x)$ by the discrete set of operators $d_{j\sigma}$,
where $a$ is the lattice spacing and $j$ the site index. This results
in a Hamiltonian of the form \eqref{eqModelChain2}. Treating the
second derivative in the kinetic term using a finite difference
method, the single-particle part of the Hamiltonian takes the form
$H_0=\sum_{ij\sigma} h^\sigma_{ij}$, with
\begin{equation}
h_{ij}^\sigma=(V_i-\tfrac{\sigma B}{2})\delta_{ij} -\tau (\delta_{i,i+1}+\delta_{i,i-1}),
\end{equation}
where $V_i$ is just the discretized version of the effective 1D potential, $B$ is the magnetic field, and $\tau=\frac{\hbar^2}{2m^*a^2}$ is the hopping matrix element. 
We define a discretized form of the interaction by
\begin{align}
 U_{ij}:=&U(a i,a j), \ \text{if} \ i\neq j \\
 U_{ii}:=&\frac{1}{a^2}\int_{a(i-1/2)}^{a(i+1/2)} dx_1 \int_{a(i-1/2)}^{a(i+1/2)} dx_2 U(x_1,x_2),
\end{align}
where we treat the on-site case separately, since $U(x_1,x_2)$ has an integrable singularity as $x_1$ approaches $x_2$.
The above treatment presupposes that the transverse wavefunctions do not change significantly on a scale set by $a$.  
If $a$ is much smaller than the characteristic length of the
electrostatic potential, the above discretization scheme correctly
captures the physical behavior of the continuous theory while
regularizing the short-distance of the interaction, with
$U_{ii}=-\frac{e^2}{\kappa\sqrt{\pi} l_y(a i)}\cdot \log[a/l_y(a
i)]+\mathcal{O}(1)$ for $a\to 0$.

Having arrived at the discretized Hamiltonian \eqref{eqModelChain2},
let us take a final look at the parameters that characterize our
system.  From the dimensionful constants $\hbar$, $e^2/\kappa$ and
$m^*$ one can construct an intrinsic length scale
$\left[\frac{\hbar^2}{m^*e^2}\kappa\right]\approx 10{\rm nm}$ and
intrinsic energy scale
$\left[\frac{m^*e^4}{2\hbar^2 \kappa^2}\right]\approx 5.5{\rm meV}$.
It is possible to express all our model's length and energy scales in
terms of these two dimensionful constants.  However, it is often
convenient to be able to relate quantities like the gate-voltage
dependence of the conductance or the spatial resolution of the density
directly to the geometry of the QPC.  For this reason, we introduce in
our studies below for each QPC a characteristic energy scale
$\bar{\Omega}_x$, and a corresponding length scale
$\bar{l}_x=\hbar/\sqrt{2 m^* \bar{\Omega}_x}$, which we measure in
absolute units and which characterize the mean geometry of the QPC
barrier. Concretely, we will take for $\bar{\Omega}_x$ the curvature
of the bare barrier at the renormalized conductance pinchoff gate
voltage $V_g^\pinchoff$, where the conductance just begins to
increase from zero (and the barrier height is
$\varepsilon_F+V_g^\pinchoff$).  All the other geometric
quantities are then specified relative to $\bar{\Omega}_x$.  To be
specific, we will characterize our QPC by the following rescaled
dimensionless quantities (denoted by tildes):
\begin{flalign}
\label{dimensionless_parameters}
&\text{(i)} \hspace{-2mm} & \widetilde{\Omega}_x&=\frac{\bar{\Omega}_x}{\rm meV} , 
&  &\text{(ii)} \hspace{-2mm} &\widetilde{V}_g&=\frac{V_g}{\bar{\Omega}_x} ,              &  &\text{(iii)} \hspace{-2mm} & \widetilde{\Omega}_y &=
\frac{\Omega_y(0)}{\bar{\Omega}_x} , & \nonumber \\ 
&\text{(iv)} \hspace{-2mm}  & \widetilde{l}_s&=\frac{l_s}{\bar{l}_x} , 
 &  & \text{(v)} \hspace{-2mm} & \widetilde{L}_{\rm bar}&=\frac{L_{\rm bar}}{\bar{l}_x} , &  &\text{(vi)}  \hspace{-2mm}  & \widetilde{x}&=\frac{x}{\bar{l}_x} ,                       & \nonumber \\
&\text{(vii)} \hspace{-2mm} 
& \widetilde{\Omega}_y''&=\frac{\bar{l}_x^2}{\bar{\Omega}_x} \left[\frac{\partial}{\partial_x^2} \Omega_y(x)\right]_{x=0}. \hspace{-4cm} & 
\end{flalign}
$\widetilde{\Omega}_x$ describes the longitudinal barrier curvature in
units of meV, $\widetilde{V}_g$ the normalized gate voltage,
$\widetilde{\Omega}_y$ the transverse curvature at the barrier center,
$\widetilde{l}_s$ the screening length, $\widetilde{L}_{\rm bar}$ the
total barrier length which controls the behavior of the flanks,
$\widetilde{x}$ the longitudinal coordinate, and
$\widetilde{\Omega}_y''$ the $x$-dependence of the transverse
curvature at the barrier center. Note that if one chooses to
specify $\widetilde{\Omega}_x$, $\widetilde{\Omega}_y$,
$\widetilde{\Omega}''_y$, $\widetilde{l}_s$, and
$ \widetilde L_{\rm bar}$, this implicitly also fixes
$\varepsilon_F$: its values has to be chosen in such a way that the
resulting curvature at pinchoff has the specified value
$\widetilde{\Omega}_x$.

It is instructive to express the interaction $U(x_1,x_2)$ of
Eq.~\eqref{interaction_1d} in terms of the rescaled dimensionless
parameters.  If we define $U_b=e^2/(\kappa \bar{l}_x)$, the
dimensionless ratio $\tilde{U}(\widetilde{x}_1,\widetilde{x}_2) =U(x_1,x_2)/U_b$
depends only on the dimensionless parameters
\eqref{dimensionless_parameters}(ii)-(vii), but not on
$\bar{\Omega}_x$. Thus, the dependence of the interaction strength (in
absolute units) on the longitudinal curvature $\bar{\Omega}_x$ of the
QPC is fully encapsulated in $U_b$.  The corresponding dimensionless
parameter
\begin{align}
 \widetilde{U}_b=U_b/\bar{\Omega}_x=\frac{\sqrt{2m^*}e^2}{\kappa \hbar} \frac{1}{\sqrt{\bar{\Omega}_x}} 
\end{align}
characterizes the effective onsite interaction strength at the barrier
center for the present long-ranged interaction model, and plays a role
analogous to the parameter $U^{\rm eff}_0=U\cdot \mathcal{A}^0_0(\mu)$
of Eq.~\eqref{Ueff} (which likewise scales as
$1/\sqrt{\bar{\Omega}_x}$) for the onsite interaction model of section
\ref{secfRGresults}. Evidently, $\widetilde{U}_b$ increases with
decreasing $\bar{\Omega}_x$, implying that interactions become ever
more important the smaller the curvature of the barrier top. Typical
values for $\widetilde{U}_b$ for the plots below range between $4.2$
and $4.9$.

\begin{figure}
 \includegraphics[width=\columnwidth]{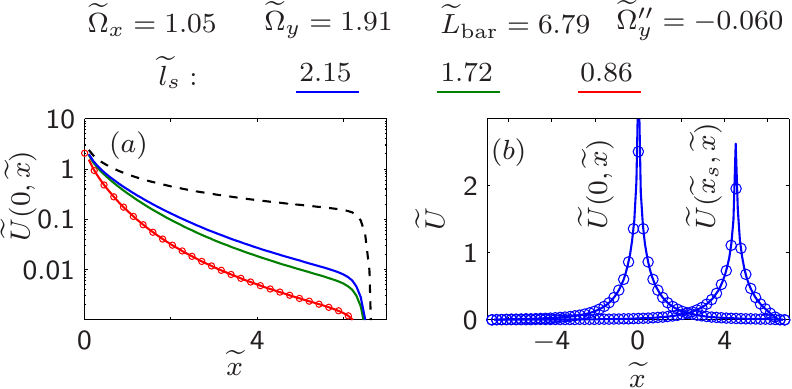}
 \caption{\label{Paper_lr_numerics_a}(a) 
   Distance dependence of the bare
   interaction $\widetilde{U}(0,\widetilde{x})$ between an electron
   located at the QPC center and one at $\widetilde{x}$, plotted on a
   logarithmic scale, for three values of $\widetilde{l}_s$.
   The dashed black line shows the
   limit of $\widetilde{l}_s \to \infty$ and the dots on the lowest curve (red)
   illustrate the chosen discretization points for the case $N=61$.
   (b) $\widetilde U(0,\widetilde x)$ (central peak) and
     $\widetilde{U}(\widetilde{x}_s=4.5 ,\widetilde{x})$ (side peak), plotted for
     $\widetilde{l}_s=2.15$ on a linear scale for both negative and
     positive $\widetilde{x}$ values.}
\end{figure}

\changed{The spatial structure of the long-ranged interaction for typical choices for
the physical parameters is shown in
Fig.~\ref{Paper_lr_numerics_a}. In Fig.\ref{Paper_lr_numerics_a}(a), we plotted the
dimensionless ratio
$\widetilde{U}(0,\tilde{x})=U(0,\widetilde{x}\cdot \bar{l}_x)/U_b$ for
three values of the rescaled screening length $\widetilde{l}_s$, as a
function of positive $\widetilde{x}=x/\bar{l}_x$. This ratio is
independent of $\bar{\Omega}_x$ itself, but increases significantly
with increasing screening length. In (b), we again show
$\tilde{U}(0,\widetilde{x})$ (central peak) and for comparison also
$\tilde{U}(\widetilde{x}_s,\widetilde{x})=U(\widetilde{x}_s\cdot
\bar{l}_x,\widetilde{x}\cdot \bar{l}_x)/\tilde{U}_b$
for fixed $\widetilde{x}_s=4.5$ as function of $\widetilde{x}$, where
the $\widetilde{x}$ range contains now the whole QPC. Due to the
reflection symmetry of our system about the QPC center,
$\widetilde{U}(0,\widetilde{x})$ is a symmetric function of
$\widetilde{x}$. In contrast,
$\widetilde{U}(\widetilde{x}_s,\widetilde{x})$ is an asymmetric
function of $\widetilde{x}$ around the point
$\widetilde{x}=\widetilde{x}_s$, decreasing more quickly when
$\widetilde{x}-\widetilde{x}_s$ becomes large positive than large
negative, because the transverse potential is wider in the former
case. This widening of the transverse potential is also the reason why
$\tilde{U}(\widetilde{x}_s,\widetilde{x})$ as a function of
$\widetilde{x}_s-\widetilde{x}$ with fixed $\widetilde{x}_s$ is in
general smaller than $\tilde{U}(0,\widetilde{x})$ as a function of
$\widetilde{x}$.}

\subsection{Discretization dependence}
\begin{figure}
 \includegraphics[width=\columnwidth]{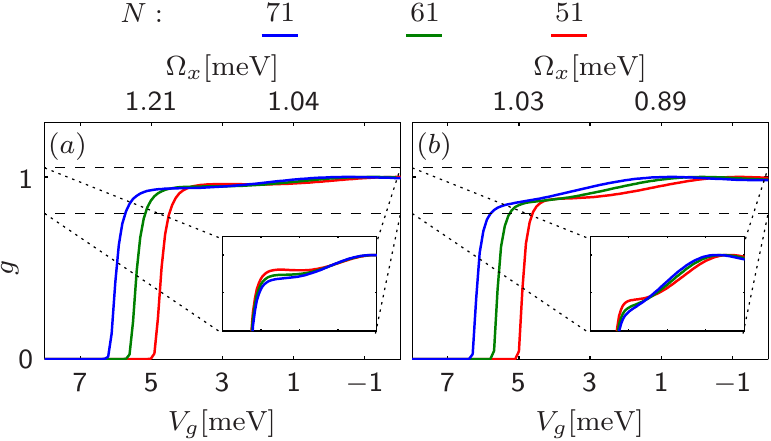}
 \caption{\label{Paper_discretization} QPC conductance
   step shape for three choices of the number of discretization points
   $N$ (with maximal feedback length $L=N-1$), for two QPCs with
   different curvatures.  We used the following parameters, in absolut
   units [c.f.\ Eqs.~\eqref{eqQPCcont} and \eqref{interaction_1d}]: In
   (a), $\gamma=0.85$, $\varepsilon_F=13.89$meV, $\Omega_y=2.35$meV,
   $L_{\rm bar}=146.11$nm, $l_s=46.17$nm; and in (b), $\gamma=0.85$,
   $\varepsilon_F=11.00$meV, $\Omega_y=2.00$meV,
   $L_{\rm bar}=158.24$nm, $l_s=50.00$nm.  The insets zoom into the
   range $g\in [0.8, 1.05]$ and plot $g$ as function of
   $V_g-V_g^\pinchoff$ to align the pinchoffs.  When expressed in
   terms of the dimensionless parameters of
   Eq.~\eqref{dimensionless_parameters}, the parameter choices in (a)
   and (b) differ only in $\widetilde{\Omega}_x$. For example, for the
   middle $N=61$ curves (green) we obtain for panel (a)
   $\mathfrak{A}=\{\widetilde{\Omega}_x=1.23,\,
   \widetilde{\Omega}_y=1.91,\, \widetilde{L}_{\rm bar}=6.79,\,
   \widetilde{\Omega}_y''=-0.060,\, \widetilde{l}_s=2.15 \}$,
   and for panel (b),
   $\mathfrak{B}=\{\widetilde{\Omega}_x=1.05,\,
   \widetilde{\Omega}_y=1.91,\, \widetilde{L}_{\rm bar}=6.79,\,
   \widetilde{\Omega}_y''=-0.060,\, \widetilde{l}_s=2.15 \}$.}
\end{figure}

We begin our treatment of long-ranged interactions by investigating to
what extent our results depend on the number of discretization points,
$N$, with all other parameters held
fixed. Fig.~\ref{Paper_discretization} shows this dependence for two
QPCs whose parameters were chosen to yield somewhat different
ranges of $\Omega_x$ curvatures. 
The first point to notice involves the $V_g$ value of the conductance
pinchoff: whereas in the absence of interactions it occurs near
$V_g=0$ , turning on our long-ranged interactions shifts it towards
the left, i.e.\ towards a larger gate voltage. This behavior is
unphysical, since for any fixed $V_g$ at which the density is nonzero,
turning on interactions should generate a Hartree barrier that causes
the conductance to decrease, not increase. We suspect that this
unphysical behavior is a fRG artefact, possibly due to our use of the
static approximation. We leave the issue of exploring what will happen
when using a dynamic version of our eCLA as a topic for future study.
We remark, however, that similar unphysical  shift artifacts where
encountered in Ref.~\onlinecite{Bauer2014} when comparing various
different fRG methods that treated the details of the vertex flow in
somewhat different ways.  Nevertheless, although the $V_g^\pinchoff$
values of the conductance curves in Ref.~\onlinecite{Bauer2014}
depended on methological details, the overall shape of the conductance
steps were essentially the same, i.e.\ when plotted as functions of
$V_g-V_g^\pinchoff$, they coincided.  We find a similar trend here: if
we increase $N$, $V_g^\pinchoff$ increases, because changing $N$
slightly changes the strength and shape of the interaction function
$U_{ij}$, causing corresponding changes in $V_g^\pinchoff$ and
$\bar{\Omega}_x$; however, the \emph{shape} of the conductance steps
in Figs.~\ref{Paper_discretization} (a),(b) seems at least
qualitatively convergent when $N$ increases [c.f.\ insets in (a) and
(b)], despite the $N$ dependence of the step's position. For the
remainder of this paper we will therefore only address the overall
shape of the conductance step.

In Fig.~\ref{Paper_discretization}(a),(b) we expressed all parameters in terms of absolute units. 
In \changed{most of} the remaining plots where physical properties are discussed
, we use
instead the more convenient dimensionless quantities introduced in
Eq.~\eqref{dimensionless_parameters} (and denoted by tildes).  We have
also extracted these dimensionless parameters for
Figs.~\ref{Paper_discretization}(a),(b) and summarized them for
further use in the parameter sets $\mathfrak{A}$ and $\mathfrak{B}$
given in the caption of Fig.~\ref{Paper_discretization}.

\begin{figure}
 \includegraphics[width=\columnwidth]{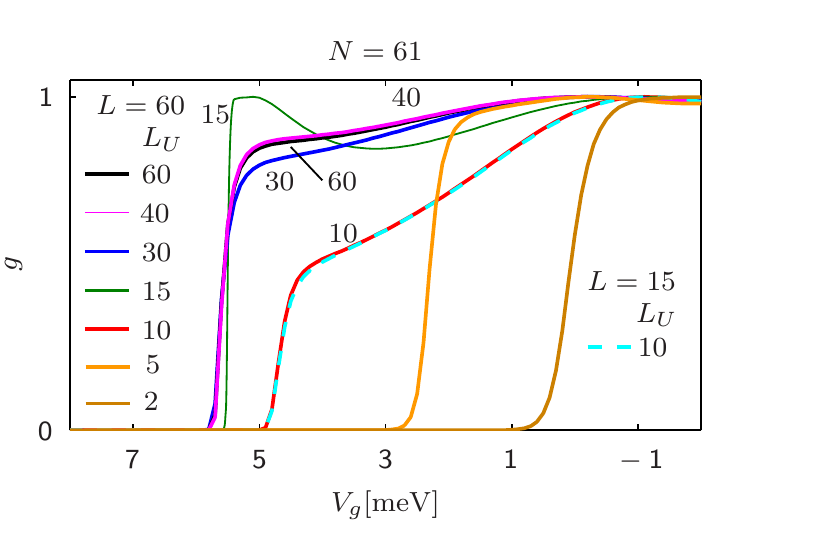}
 \caption{\label{Paper_vary_Lu}QPC conductance curves
   at fixed $N$, calculated with feedback length $L=N-1$ for several
   values of the interaction cutoff $L_U$ (solid lines), and with
   $L=15$ for $L_U=10$ (dashed line). The QPC parameters were chosen
   as in Fig.~\ref{Paper_discretization}(b). Note that while
   convergence in $L$ is rapid, the conductance becomes independent of
   the cutoff length only for $L_U>40$.\changed{ Furthermore, for $L_U \lesssim l_x/a\approx 4.4$ we recover the conductance shape of short-ranged interactions.}}
\end{figure}

In Fig.~\ref{Paper_discretization} we used the maximal feedback length
$L=N-1$ to fully take interactions over the whole QPC into
account. However, due to numerical costs, this limited the number of
sites that could be treated to $N\leq 71$.  For this reason, we have
also explored using a cutoff length $L_U$ for the interaction range,
setting $U_{ij}=0$ for $|i-j|>L_U$. \changed{The resulting conductance
  curves for different $L_U$ are shown in Fig.~\ref{Paper_vary_Lu}.}
\changed{We first note that when the cutoff length $L_U$ becomes smaller
  than the characteristic length $l_x/a \approx 4.4$ of the QPC, we
  recover the conductance shape for short-ranged interactions. This
  behavior is analogous to that obtained in
  Fig.~\ref{Paper_lr_numerics_b} below, when reducing the screening
  length $l_s$ below $l_x$.}  Furthermore, we find rapid convergence
when increasing $L$ beyond $L_U$ for a fixed $N$: for example,
Fig.~\ref{Paper_vary_Lu} contains two curves for $L_U=10$, one
computed with $L=60$ (solid), the other with $L=15$ (dashed), which
essentially coincide.  However, the shape of the conductance step
becomes independent of $L_U$ only for rather large values of $L_U$,
implying that the tail of the long-ranged interaction actually matters
significantly. Therefore, we did not pursue using $L_U < N$ any
further and for the remainder of this work show only data obtained
without interaction cutoff and with full feedback length, $L=N-1$.
 
\subsection{Effects of long-ranged interactions on QPC properties}
After these technical considerations, let us now study how the
fact that the interaction range is not zero affects the QPC
properties. For this, we first briefly discuss the dependence of our
finite-ranged interaction on the given physical parameters and then
study the resulting consequences on the conductance and the density.
As pointed out earlier, this study does not aim to achieve a
fully realistic description of screening in a QPC, but rather serves
as a first illustration of the potential of the eCLA for treating a
model with reasonably long-ranged interactions.

\begin{figure}
 \includegraphics[width=\columnwidth]{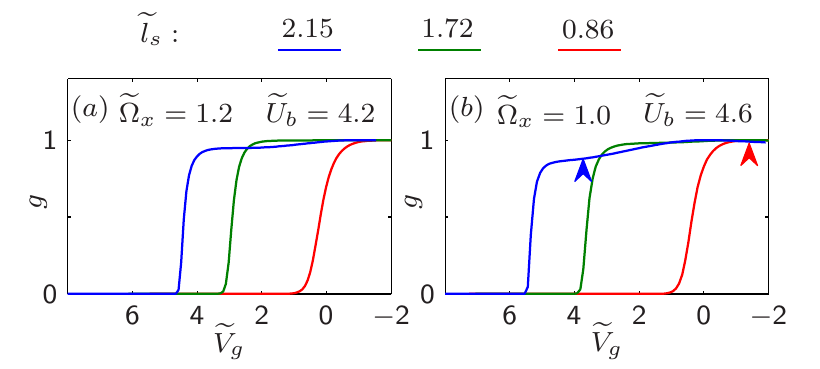}
 \caption{\label{Paper_lr_numerics_b}\changed{(a), (b) The conductance curves corresponding to the interactions depicted in Fig.~\ref{Paper_lr_numerics_a} (a), for two
   different QPC mean curvatures $\widetilde{\Omega}_x=1.2$ and $\widetilde{\Omega}_x=1.0$,
   respectively. The arrows at the right (red) $\widetilde{l}_s=0.86$ and the
   left (blue) $\widetilde{l}_s=2.15$ curve in (b) indicate the gate voltages
   $\widetilde{V}_g=-1.43$ and $\widetilde{V}_g=3.73$ at which the
   density profiles in Figs.~\ref{Paper_dichte_osc}(a) and (b) were
   calculated, respectively.}}
\end{figure}

Figs.~\ref{Paper_lr_numerics_b}\changed{(a) and (b)} show, for two different
values of the curvature $\widetilde{\Omega}_x$, respectively, three
conductance curves corresponding to the three choices of
$\widetilde{l}_s$ used in Fig.~\ref{Paper_lr_numerics_a}(a). For both
choices of $\widetilde{\Omega}_x$, we obtain an onsite-like
conductance step shape when $\widetilde{l}_s$ is small. When
$\widetilde{l}_s$ is increased, i.e.\ when the amount of screening
is reduced, the step shape acquires some additional features, such
as the emergence of a ``preplateau'' at a value of $g$ slightly
lower than 1, followed by a much slower increase towards $1$ in 
Fig.~\ref{Paper_lr_numerics_b}(a).  These features are more pronounced
for the longer QPC (i.e.\ smaller curvature) of
Fig.~\ref{Paper_lr_numerics_b}(b), where the conductance quickly
reaches a preplateau around $g\simeq 0.8$ and thereafter increases much
more slowly.  

\begin{figure}
 \includegraphics[width=\columnwidth]{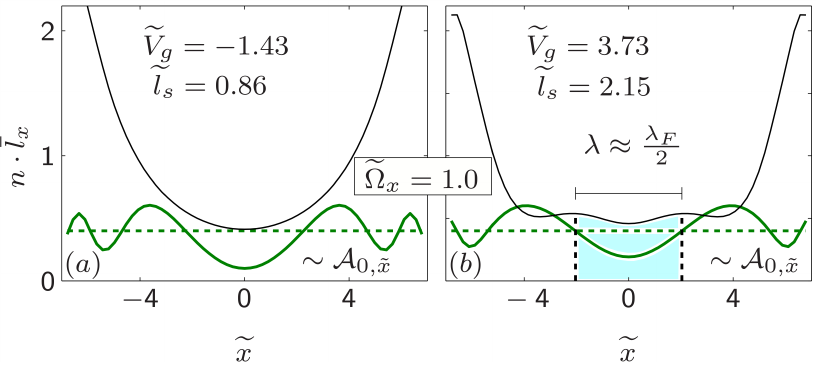}
 \caption{\label{Paper_dichte_osc}Density profiles
   (thin lines) calculated for two fixed parameter choices from
   Fig.~\ref{Paper_lr_numerics_b}, indicated for panels (a) and (b)
     by the right and left arrows in
     Fig.~\ref{Paper_lr_numerics_b}(b), respectively. For comparison,
     the thick lines depict (a vertically rescaled version of) the imaginary part of the
     interacting single-particle propagator at the chemical potential,
     $\mathcal{A}_{0,\widetilde{x}}=-\frac{1}{\pi
       \bar{l}_x}\operatorname{Im} G^R_{0,\widetilde{x}}(\omega=0)$.
     Horizontal dashed lines indicate where
     $\mathcal{A}_{0,\widetilde{x}}=0$.  In (b), the distance between
     the two density maxima (marked by the dashed vertical lines) is
     $\lambda=3.62\bar{l}_x$. This agrees well with two estimates of
     $\lambda_F/2$, either from the distance between the two central
     zeros of $\mathcal{A}_{0,\widetilde{x}}$ finding
     $\lambda_F/2=3.82\bar{l}_x$ or from the mean density $\bar{n}$ in
     the center of the QPC (shaded region) finding
     $\lambda_F/2=3.55\bar{l}_x$.}
\end{figure}

In order to explore the origin of this behavior, we show in
Fig.~\ref{Paper_dichte_osc}(a) and \ref{Paper_dichte_osc}(b) two
density profiles (thin lines), calculated, respectively, for two
fixed parameter choices from Fig.~\ref{Paper_lr_numerics_b}(b),
indicated in the latter by the right (red) marker for
$\widetilde{l}_s\!=\!0.86$, $\widetilde{V}_g\!=\!-1.43$ and the left
(blue) marker for $\widetilde{l}_s\!=\!2.15$,
$\widetilde{V}_g\!=\!3.73$. In Fig.~\ref{Paper_dichte_osc}(b), for
which the rescaled screening length $\widetilde{l}_s$ is larger, we
observe three qualitative changes relative to
Fig.~\ref{Paper_dichte_osc}(a). First, the flanks of the density
profile are somewhat steeper. Second, the spatial region in which
the density is low has become wider. And third, in this
low-density region the density shows some weak density
oscillations that are absent in Fig.~\ref{Paper_dichte_osc}(a). 

The first two features suggest that the long-range interactions
have generated a renormalized barrier whose shape has a flatter
top and steeper flanks than the bare parabolic barrier.  This
flattening occurs because the bare density is larger in the flanks
than near the center, hence the upward Hartree-type shift of the
barrier potential, which is proportional to the bare density, is
larger in the flanks than near the center. The upward
renormalization in the flanks becomes stronger the larger the
interaction range, because then the upward Hartree-type shift at a
given site is determined by a weighted average of the density over
a range of nearby sites (whose extent is set by the screening
length), and since the bare density profile is convex, the sites
in the flanks contribute more strongly.

To shed further light on the third feature, namely the weak density
oscillations in the low-density region, we compare their oscillation period 
 with estimates for the ``local Fermi wavelength''
$\lambda_F$ at the QPC center, which can be extracted from either the
interacting Green's function or the mean density in the center of the
QPC. To illustrate the first method, the thick lines in
Figs.~\ref{Paper_dichte_osc}(a) and (b) indicate the oscillatory
behavior of
$\mathcal{A}_{0,\widetilde{x}}=-\frac{1}{\pi
  \bar{l}_x}\operatorname{Im} G^R_{0, \widetilde{x}}(\omega=0)$.
For a homogeneous system the Green's function oscillates with period
$\lambda_F$, and likewise we can here define an effective
$\lambda_F/2$ in the middle of the QPC by taking the distance between
the two central zeros of the thick line. For
Fig.~\ref{Paper_dichte_osc}(b), the position of these zeros is in good
agreement with the position of the density maxima of the QPC
(indicated by the two dashed vertical lines), whereas the density in
Fig.~\ref{Paper_dichte_osc}(a) shows no features on the scale of
$\lambda_F$.  An alternative way to extract an effective $\lambda_F$
is to calculate the mean density $\bar{n}$ in the center of the QPC
between the two density maxima (shaded region in
Fig.~\ref{Paper_dichte_osc}), and use
$\lambda_F=2\pi/k_F=4/\bar{n}$. For Fig.~\ref{Paper_dichte_osc}(b),
the first method yields $\lambda_F/2=3.82\bar{l}_x$, and the second
$\lambda_F/2=3.55\bar{l}_x$, which are both in reasonable agreement
with each other and the distance $\lambda=3.62\bar{l}_x$ between the
two density maxima. Thus, we conclude that the period of the density
oscillations observed here can be associated with $\lambda_F/2$,
or equivalently wavenumber $2k_F$.

\begin{figure}
\includegraphics[width=\columnwidth]{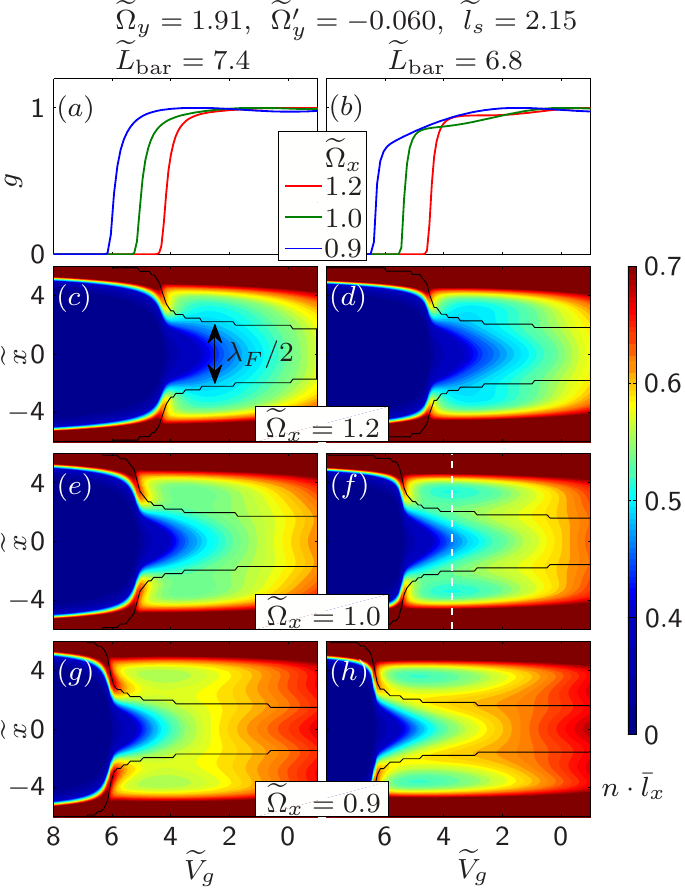}
\caption{\label{Paper_lr_vary_lx}Study of two QPCs
  with different $\widetilde{L}_{\rm bar}$, for three choices of
  $\widetilde{\Omega}_x$. The other dimensionless parameters were
  chosen the same as in $\mathfrak{B}$ [c.f.\ caption of
  Fig.~\ref{Paper_discretization}]. (a,b) Conductance as function of
  gate voltage, and (c-h) density as function of position and gate
  voltage.  While the conductance changes its shape for
  both QPCs, the shorter one (b) shows stronger features, preeminently
  a shoulder in the conductance step. In the density, both QPCs show
  the development of oscillations with approximate wavelength
  $\lambda_F/2$, which is determined by the Green's function as in
  Fig.~\ref{Paper_dichte_osc} and indicated by the distance between
  the black lines. In the last plots (g) and (h) the density
  oscillations transition at smaller gate voltages from two to three
  maxima. The cut along the dashed white line in (f) is precisely the
  density profile plotted in Fig.~\ref{Paper_dichte_osc}(b).}
\end{figure}

In Fig.~\ref{Paper_lr_vary_lx} we examine this behavior more systematically, using two QPCs having a comparatively long screening length of $\widetilde{l}_s=2.15$, but which differ slightly in $\widetilde{L}_{\rm bar}$, i.e.\ in their total barrier length.
For both QPCs the conductance step
[Figs.~\ref{Paper_lr_vary_lx}(a),(b)] changes its shape with
decreasing curvature $\widetilde{\Omega}_x$ and for the right QPC with
smaller $\widetilde{L}_{\rm bar}$ develops additional pronounced
features in the plateau region. In Figs.~\ref{Paper_lr_vary_lx}(c)-(h)
we show the corresponding densities (color scale) as functions of gate
voltage and longitudinal position, and find that with decreasing
curvature $\widetilde{\Omega}_x$ the density develops
oscillations. The period of these oscillations is again set by
$\lambda_F/2$, which is indicated in
Figs.~\ref{Paper_lr_vary_lx}(c)-(h) by the distance between the black
lines. While for the right QPC the two density maxima follow very
accuratly the black lines, in the left QPC they lie slightly further
apart than $\lambda_F/2$. The reason for this might be that the left
QPC is slightly longer ($\widetilde{L}_{\rm bar}$ is larger), giving
the electrons in the center more space to form the two repelling
density maxima, but not enough space to fit a third density maximum
into the available region.  In summary, we find that when increasing
the geometric proportions of the QPC compared to the scale set by the
interactions, i.e.\ when decreasing
$\widetilde{\Omega}_x$, the conductance develops additional features
in the plateau region, and simultaneously density oscillations arise
on a scale set by $\lambda_F/2$. 
 
\begin{figure}
 \includegraphics[width=\columnwidth]{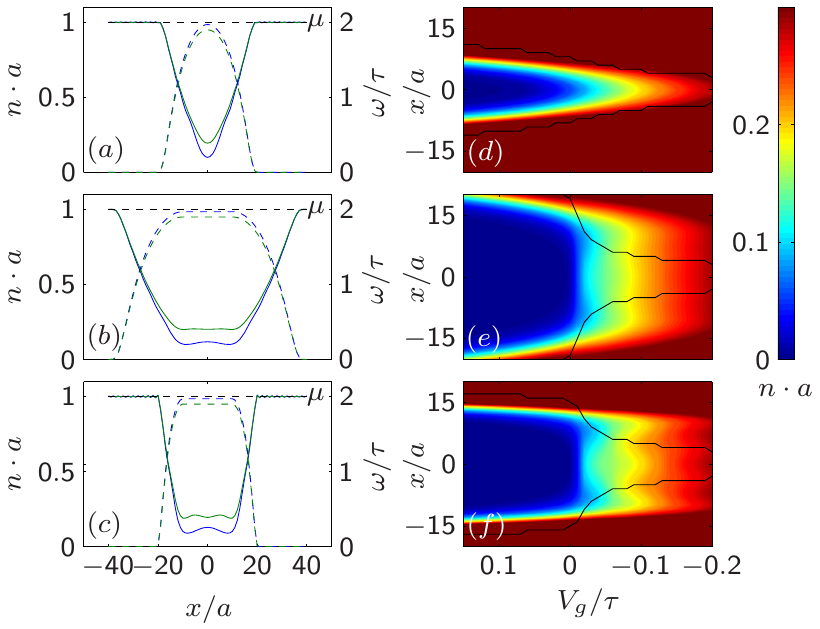}
 \caption{\label{fig:bare-barrier}(a-c) Barrier
   shapes (dashed lines) and corresponding noninteracting densities
   (solid lines) for almost open QPCs with (a) a parabolic barrier
   top, (b) a flat barrier top with wide flanks, (c) and a flat
   barrier top with steep flanks. (d-f) Density profiles corresponding
   to these three barrier shapes, plotted as functions of position and
   gate voltage.  In these plots, $\lambda_F/2$ is again indicated by
   the distance between the black lines. The flat barrier top with
   steep flanks of panel (c) yields pronounced Friedel oscillations in
   the density profile shown in panel (f), which resemble the density
   oscillations caused by the long-range interaction in the open
   regime of the QPCs of Fig.~\ref{Paper_lr_vary_lx}(e-h). This
   suggests that for the latter, the renormalized barriers have a
   rather flat tops with steep flanks. }
\end{figure}

We interpret the $2 k_F$ density oscillations seen in
Fig.~\ref{Paper_dichte_osc}(b) as Friedel oscillations generated by
the inhomogeneity induced by the renormalized QPC potential.  A
similar interpretation was envoked in Iqbal \textit{et al.}
\cite{Iqbal2013} where they also found a wavelength $\lambda_F/2$,
or equivalently a wavenumber of $2 k_F$, for their spin polarized,
emergent localized states (ELS) obtained from SDFT calculations in
long QPCs.

To support this interpretation, we show in 
\mbox{Figs.~\ref{fig:bare-barrier}(a-c)}
some density profiles (solid lines) obtained for a QPC model of
\textit{noninteracting} electrons traversing a QPC, comparing three
different barrier shapes (dashed lines): (a) a parabolic top, (b) a
flat top with a slow transition to broad flanks, and (c) a flat
top with a rather quick transition to steep flanks. For a given
gate voltage, the overall shape of the density profile mirrors that
of the barrier top for all three cases. Moreover, pronounced additional
density oscillations arise for case (c). 
Panels (d) to (f) show the corresponding evolution of such density
profiles with gate voltage.  For gate voltages where the QPC is
sufficiently open that the density in the center is not very low,
the density oscillations seen in Figs.~\ref{fig:bare-barrier}(c) and
\ref{fig:bare-barrier}(f) are reminiscent, respectively, of those
seen in Figs.~\ref{Paper_dichte_osc}(b) and
\ref{Paper_lr_vary_lx}(c-h) for QPCs with interactions whose range
is longer than the characteristic QPC length (i.e.\ with
$\widetilde l_s > 1$). This supports the interpretation offered
above that such QPCs indeed have renormalized barriers with rather
flat tops and steep flanks.  However, for higher gate voltages where
the QPC is beginning to close off and the density in the center
becomes very low, we see a qualitative difference between the
density profiles shown in Fig.~\ref{fig:bare-barrier}(f)  and those
of Figs.~\ref{Paper_lr_vary_lx}(c-h): the former shows a weak
density maximum, whereas the latter do not, because in the regime of
very low densities, the Hartree-type renormalization of the barrier
shape is not yet strong enough to generate a flattish barrier top.

\begin{figure}
 \includegraphics[width=\columnwidth]{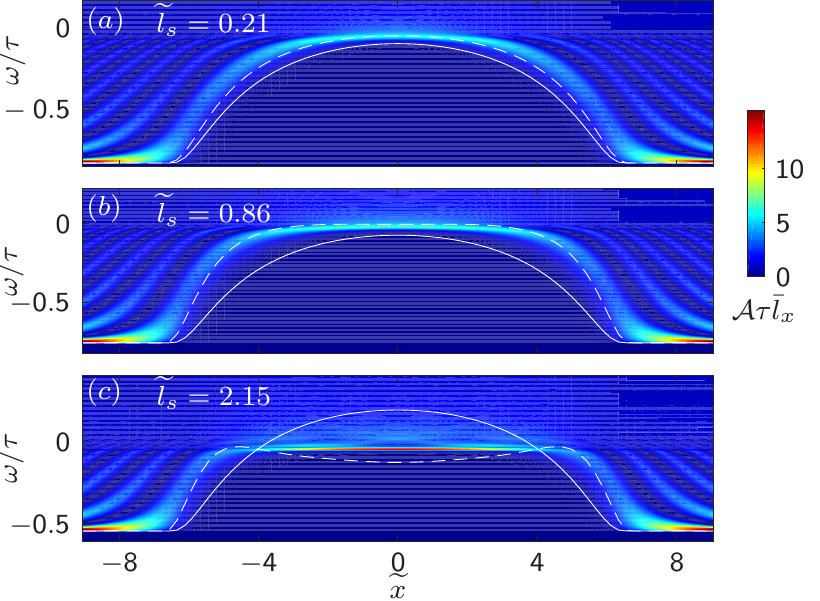}
 \caption{\label{fig:flattend_ldos}\changed{Interacting
     LDOS in the static approximation [Eq.~\eqref{eq:intldos}], shown
     as function of position and energy (color scale), for three
     different values of the screening length $\widetilde{l}_s$.
     Solid white lines show the bare potential $V_j$, and dashed white
     lines $V_j + \Sigma_{jj}$, as functions of position. The physical
     parameters used for this plot correspond to those of
     Fig.~\ref{Paper_lr_numerics_b}(b), with the gate voltage was set
     to $\widetilde{V}_g=-1.91$ in (a), $\widetilde{V}_g=-1.43$ in (b)
     and $\widetilde{V}_g=3.73$ in (c). (The latter two correspond to
     the red and blue markers in Fig.~\ref{Paper_lr_numerics_b}.) The
     shape of the band bottom reflects that of the renormalized
     barrier.  (The fact that the renormalized barrier top lies
     \emph{below} the bare barrier top in (c) is due to the artifact
     of static fRG discussed in section IV.B.) }  \vspace{-0.5cm}}
\end{figure}

\changed{To further explore our hypothesis concerning the occurence of a
  renormalized barrier with a flattened top and steep flanks, we have studied the influence of the
  screening length, $\widetilde{l}_s$, on the interacting LDOS in the
  static approximation,
\begin{align}
\mathcal{A}_j(\omega)&=-\tfrac{1}{\pi}\operatorname{Im} G_{jj}(\omega+ i0^+)
=-\tfrac{1}{\pi}\operatorname{Im}[\omega - h^0 - \Sigma]^{-1}_{jj},
\label{eq:intldos}
\end{align}
where
$h^0_{ij}= \delta_{ij} V_j - \tau [\delta_{i,j+1} + \delta_{i,j-1}]$
is the bare single-particle Hamiltonian,
and $\Sigma_{ij}$ is the static self-energy at the end of the RG flow
\cite{need-dynamic-fRG}. Fig.~\ref{fig:flattend_ldos} shows the LDOS
(color scale) as function of position and energy, for three values of
the screening length, $\widetilde{l}_s$.}  \changed{We interpret the shape
of the effective band bottom as indicative of the shape of the
effective barrier.  We observe that with increasing $\widetilde{l}_s$,
the effective barrier top indeed does become strikingly flat over
an extended region of space centered on the middle of the QPC, ending
in rather steep flanks, as anticipated above.  For comparison, solid
white lines show the bare potential $V_j$ with its parabolic
top. Moreover, dashed white lines show $V_j+\Sigma_{jj}$, to
illustrate the contribution of the diagonal elements of the
self-energy to the renormalization of the potentatial barrier.
However, while $V_j+ \Sigma_{jj}$ does show a trend toward barrier
flattening with increasing screening length, for the largest
$\widetilde{l}_s$ value [Fig.~\ref{fig:flattend_ldos}(c)] it leads to
a shallow local minimum at $\widetilde{x}=0$, reminiscent of a QD-like
barrier shape.
To correctly capture the shape of the band bottom, which shows no such
local minimum, the off-diagonal elements of the self-energy have to be
taken into account, too. This is done when computing the LDOS according to
Eq.~\eqref{eq:intldos}, which involves inverting the entire
matrix $\omega - h^0 - \Sigma$ before taking diagonal elements.

The above results show that long-range interactions can have a rather
striking flattening effect on the effective barrier shape, and that
long flat barriers lead to interesting density oscillations.  It would
thus} be interesting to study the geometric crossover from a QPC to a
homogeneous wire obtained by making the QPC length $\bar l_x$ very
long, or by using flat-topped bare barriers of increasing width. In a
paper by Schulz \cite{Schulz1993}, concerning Wigner crystal physics
in 1D, it was predicted that in a homogeneous 1D model with
long-ranged Coulomb interactions in the low-density limit, the
density-density correlator $\langle \rho (x) \rho(0) \rangle$ contains
both $2 k_F$ and $4 k_F$ oscillations.  The latter decay more slowly
with $x$, and are argued by Schulz to lead to a Wigner crystal in a
homogeneous system.  During the aforementioned geometric crossover
from a QPC to a long wire, well-developed $4k_F$ density oscillations
can be expected to emerge, which could be regarded as precursors for
the formation of a Wigner crystal.  A systematic study of this
behavior would be extremely interesting, but falls beyond the scope of
this paper and is left for future study. In particular, future work
would have to incorporate screening also due to higher transport
channels, leading to a shorter-ranged interaction, so that the effects
discussed above would likely turn out to be somewhat less pronounced
than found here.

\section{Conclusion and Outlook}

Building on previous works \cite{Bauer2013,Bauer2014}, we have
introduced an improved approximation scheme for 3rd-order truncated
fRG. We use an extended coupled ladder approximation (eCLA), splitting
the fRG-flow into three channels depending on the internal index
structure. When treated independently, each of these channels behaves
as in the random phase approximation.  The complexity of the
eCLA scheme depends on the amount of feedback admitted between the
individual channels. For the frequency dependence, we only used static
feedback between the channels. In order to control the amount of
feedback in the spatial structure, we have introduced the feedback
length $L$. In the case $L=0$ we get the minimal feedback between the
channels, corresponding to the CLA of previous works \cite{Bauer2014},
whereas for $L\rightarrow N-1$ we recover the full spatial vertex flow
in 2nd order.

For actual computations, we restricted ourselves to static fRG, i.e.\ in addition to using only a static feedback between the channels we also neglected the frequency dependence of the vertices altogether. In this additional approximation, we calculated the zero-temperature Green's function at the chemical potential, which is the relevant quantity in order to compute the linear conductance of the system.

We first applied our new method to a QPC model with onsite
interactions, which has extensively been studied in the past. Here, we
observed that the longer-ranged feedback leads to a quantitative but
not qualitative change as long as both methods are convergent for the
respective parameters. In particular, we observed \changed{for barriers with characteristic lengths between $4-10$ sites that convergence in $L$ is achieved onces $L$ becomes comparable to $l_x$.}  Additionally, we observed that the enhanced
feedback stabilizes the fRG-flow and therefore leads also to
convergence in parameter regimes which could not be studied with the
$L=0$ method. To illustrate this increased stability, we studied
QPC-QD crossovers analogous to those discussed by Heyder \textit{et
  al.} in Ref.~\cite{Heyder2015} using the CLA. There, the convergence
of the fRG flow suffers especially from the high LDOS at the chemical
potential that occurs during the crossover when the barrier top
becomes flat in an extended region close to the chemical
potential. Our stabilized flow, however, enabled us to study this type
of transition. In particular, we succeeded to study regimes of very
shallow dots, containing only a few electrons, and observed the Kondo
plateau in the conductance expected for such dots.

Finally, in order to test the full potential of our improved feedback, we applied it to a QPC with finite-ranged interactions. 
The most striking observation was that for a relatively flat QPC in
the regime of low density and sufficiently long-ranged interactions,
the conductance reaches a preplateau somewhat \textit{below} $g = 1$
(before slowly climbing towards $g=1$), accompanied by the onset of
oscillations in the density.  The wavelength of these density
oscillations was determined to be approximately $\lambda_F/2$,
admitting an interpretation as Friedel oscillations arising from a
renormalized barrier shape with a rather flat top and steep flanks.
This behavior is consistent with that observed by Iqbal \textit{et
  al.}  \cite{Iqbal2013} in SDFT calculations for their emergent
localized states (ELS) in a spin-polarized QPC.
It would be of great interest to explore these type of effects more
systematically in the future, within a more realistic model that
incorporates the effects of higher transport modes when deriving the
effective screened interaction for the lowest-lying transport
mode. In particular, the geometric crossover between a
QPC potential and a homogeneous quantum wire, expected to 
show Wigner crystallization, could be explored in this fashion.
However, it remains to be seen whether fRG 
will be able to cope with the truly homogeneous limit;
such a study will presumably also have to  employ 
tools more powerful than fRG, such as the density matrix renormalization group.

By way of an outlook to future technical fRG developments, let us remark
that it would be desirable to find ways of avoiding an fRG artifact
that is present in our results: upon turning on a long-ranged
interaction, the position of the conductance step shifts not to
smaller gate voltages, as physically expected, but to larger ones. We
suspect that this is artefact results from our use of static fRG. A
next possible step to remedy this problem could be to change from
static to dynamic fRG, i.e.\ to implement the frequency dependence of
the vertices.  Moreover, it would also be possible to use our enhanced
feedback scheme in the context of Keldysh fRG, which is additionally
able to treat the temperature dependence and non-equilibrium behavior
of QPCs. This would be numerically challenging since the Keldysh
scheme in the $L=0$ implementation is already very costly by
itself. However, one might profit from the fact that the most
expensive part of the Keldysh calculation scales with
$\mathcal{O}(L^2)$, and not with $\mathcal{O}(L^3)$ as in our case.
Work in that direction is currently in progress.

\begin{acknowledgments}
  We thank Jan Heyder, Volker Meden, Yigal Meir and Dennis Schimmel
  for very helpful discussions. We acknowledge support from the DFG
  via SFB-631, SFB-TR12, De730/4-3, and the Cluster of Excellence
  \emph{Nanosystems Initiative Munich}.
\end{acknowledgments}
\bibliography{tmp}
\end{document}